\documentclass[12pt,a4paper,english]{article}
\usepackage[T1]{fontenc}
\usepackage[utf8]{inputenc}
\usepackage{amsmath}
\usepackage{amssymb}
\usepackage{graphicx}
\usepackage{esint}

\makeatletter

\providecommand{\tabularnewline}{\\}

\usepackage{a4wide}
\usepackage{psfrag}
\makeatletter
\renewcommand{\theequation}{\hbox{\normalsize\arabic{section}.\arabic{equation}}}
\@addtoreset{equation}{section}
\renewcommand{\thefigure}{\hbox{\normalsize\arabic{section}.\arabic{figure}}}
\@addtoreset{figure}{section}
\renewcommand{\thetable}{\hbox{\normalsize\arabic{section}.\arabic{table}}}
\@addtoreset{table}{section}
\makeatother

\usepackage{ifpdf}\ifpdf\usepackage{epstopdf}\usepackage[pdftex,ps2pdf,dvips,colorlinks,urlcolor=blue,citecolor=blue,linkcolor=blue]{hyperref}\else
\usepackage[hypertex,ps2pdf,dvips,colorlinks,urlcolor=blue,citecolor=blue,linkcolor=blue]{hyperref}\fi\pdfadjustspacing=1

\makeatother

\usepackage{babel}
\begin{document}

\title{\begin{flushright}{\normalsize ITP-Budapest Report No. 653}\end{flushright}\vspace{1cm}Determining
matrix elements and resonance widths from finite volume: the dangerous
$\mu$-terms}

\author{G. Takács\\
 HAS Theoretical Physics Research Group\\
 1117 Budapest, Pázmány Péter sétány 1/A, Hungary}

\date{9th October 2011}
\maketitle
\begin{abstract}
The standard numerical approach to determining matrix elements of
local operators and width of resonances uses the finite volume dependence
of energy levels and matrix elements. Finite size corrections that
decay exponentially in the volume are usually neglected or taken into
account using perturbation expansion in effective field theory. Using
two-dimensional sine-Gordon field theory as ``toy model'' it is
shown that some exponential finite size effects could be much larger
than previously thought, potentially spoiling the determination of
matrix elements in frameworks such as lattice QCD. The particular
class of finite size corrections considered here are $\mu$-terms
arising from bound state poles in the scattering amplitudes. In sine-Gordon
model, these can be explicitly evaluated and shown to explain the
observed discrepancies to high precision. It is argued that the effects
observed are not special to the two-dimensional setting, but rather
depend on general field theoretic features that are common with models
relevant for particle physics. It is important to understand these
finite size corrections as they present a potentially dangerous source
of systematic errors for the determination of matrix elements and
resonance widths.
\end{abstract}

\section{Introduction}

The matrix elements of local operators (form factors) are central
objects in quantum field theory. Indeed they are the subject of considerable
interest in lattice QCD, where -- among other applications -- they
are relevant in describing the weak decays of hadrons. Due to the
Maiani-Testa no-go theorem \cite{Maiani:1990ca}, it is necessary
to extract them using finite-size methods, which are the subject of
a seminal paper by Lellouch and Lüscher \cite{Lellouch:2000pv} (see
also \cite{Lin:2001ek}). A closely related problem is description
of resonances, whose decay width can also be extracted from finite-size
data, for which the first proposal was made by Lüscher \cite{Luscher:1991cf}.

Two-dimensional integrable quantum field theories provide an ideal
testing ground for these ideas because in many such models the exact
analytic expressions of the form factors are known. First, the $S$
matrix can be obtained exactly in the framework of factorized scattering
developed in \cite{zam-zam} (for a later review see \cite{Mussardo:1992uc}).
It was shown in \cite{Karowski:1978vz} that it is possible to obtain
a set of equations satisfied by the form factors using the exactly
known scattering amplitudes as input. The complete system of form
factor equations, which provides the basis for a programmatic approach
(the so-called form factor bootstrap) was proposed in \cite{Kirillov:1987jp}.
For a detailed and thorough exposition of the subject we refer to
\cite{Smirnov:1992vz}. 

On the other hand, for two-dimensional theories there is a very efficient
alternative to lattice field theory to evaluate finite size spectra
and matrix elements called the truncated conformal space approach,
which was first proposed by Yurov and Zamolodchikov \cite{Yurov:1989yu}.

Recently we used this framework to perform a detailed analysis of
resonances and local operator matrix elements. Resonances were studied
in \cite{Pozsgay:2006wb}, where it was argued that they can be extracted
with a much better accuracy from level splittings than from the slope
method originally proposed in \cite{Luscher:1991cf}. It was realized
that the proposed method (dubbed the ``improved mini-Hamiltonian''
method) is essentially equivalent to determining the matrix element
of the interaction term responsible for the decay, and in \cite{Pozsgay:2007kn,Pozsgay:2007gx}
we explored an approach to treat general matrix elements of local
operators, which is an extension of the Lellouch-Lüscher approach.
A proof given in \cite{Pozsgay:2007kn} shows that results obtained
in this framework are valid to all orders in the inverse volume $1/L,$
i.e. up to corrections decaying exponentially with the volume. We
also realized that these so-called ``residual'' finite size corrections
could play an important role, but the tools to study a particular
class of these, the so-called $\mu$-terms were only developed later
by Pozsgay in \cite{Pozsgay:2008bf}.

In this work the issue of $\mu$-terms is addressed in more details.
The testing ground is a specific version of the well-known sine-Gordon
theory, the so-called $k$-folded model \cite{Bajnok:2000wm}, mainly
because all the necessary techniques are well-developed and the model
is thoroughly understood from the theoretical point of view. This
work can also be considered as a refinement of the form factor study
performed in \cite{Feher:2011aa}, where the relevance of $\mu$-terms
was pointed out, but their detailed analysis was omitted.

The main goal of this paper is to present the issues in a way that
leads to conclusions which are expected to be relevant to a wide class
of models, and in particular to the ongoing efforts to determine resonance
parameters and decay matrix elements in lattice QCD. 

The outline of the paper is as follows. In section \ref{sec:The-two-folded-sine-Gordon}
the necessary details concerning the two-folded sine-Gordon model
are introduced. Section \ref{sec:Form-factors-in} presents the formalism
of finite size form factors, first the description valid to all orders
in $1/L$ and then determining the leading exponential corrections
a.k.a. the $\mu$-terms. The theoretical results presented here are
compared to numerical data extracted from TCSA in section \ref{sec:Determining-matrix-elements}.
Section \ref{sec:Decay-widths-of} discusses the relevance of these
findings to the description of resonances, and in section \ref{sec:Conclusions-and-outlook}
the conclusions are formulated. The detailed and rather bulky formulae
for the exact form factors (determined from the bootstrap) are summarized
in Appendix \ref{sec:Sine-Gordon-breather-form-factors}.

\section{The two-folded sine-Gordon model in finite volume\label{sec:The-two-folded-sine-Gordon}}

The classical action of sine-Gordon theory is 
\begin{equation}
\mathcal{A}=\int d^{2}x\left(\frac{1}{2}\partial_{\mu}\Phi\partial^{\mu}\Phi+\frac{m_{0}^{2}}{\beta^{2}}\cos\beta\Phi\right)\label{eq:sg_action}
\end{equation}
A useful representation of the model at quantum level is to consider
it as the conformal field theory of a free massless boson perturbed
by a relevant operator. In this framework, the Hamiltonian can be
written as
\begin{equation}
H=\int dx\frac{1}{2}:\left(\partial_{t}\Phi\right)^{2}+\left(\partial_{x}\Phi\right)^{2}:-\lambda\int dx:\cos\beta\Phi:\label{eq:pcft_action}
\end{equation}
where the semicolon denotes normal ordering in terms of the modes
of the $\lambda=0$ massless field. Due to anomalous dimension of
the normal ordered cosine operator, the coefficient $\lambda$ has
dimension
\[
\lambda\sim\left[\mbox{mass}\right]^{2-\beta^{2}/4\pi}
\]
and it sets the mass scale of the model. The genuine coupling constant
is $\beta$ which for later convenience is reparametrized as
\[
\xi=\frac{\beta^{2}}{8\pi-\beta^{2}}
\]
The spectrum consists of a doublet of solitons and their bound states,
which are called breathers. The scattering amplitudes of this model
are briefly reviewed in Appendix \ref{sub:Spectrum-and-}. As usual
in two-dimensional kinematics, the on-shell energy-momentum two-vector
$(E,p)$ of a particle with mass $m$ is parametrized with the rapidity
$\theta$:
\[
E=m\cosh\theta\quad,\quad p=m\sinh\theta
\]
Lorentz boosts correspond to shifting all rapidities by the same constant.

In finite volume, it is possible to choose quasi-periodic boundary
conditions for the field leading to a set of possible local theories
labeled by an integer $k$ called the folding number \cite{Bajnok:2000wm}.
To define the \ensuremath{k}
-folded theory \ensuremath{SG(\beta,k)}
 we take the sine-Gordon field $\Phi$ as an angular variable with
the period
\begin{equation}
\Phi\,\sim\,\Phi+\frac{2\pi}{\beta}k\label{period}
\end{equation}
When the space is a finite circle with circumference (in this case
also the volume) $L$, i.e. $x\sim x+L$, this implies the following
quasi-periodic boundary condition for the field
\begin{equation}
\Phi(x+L,t)=\Phi(x,t)+\frac{2\pi}{\beta}km\:,\: m\in{\mathbb{Z}}\label{quasi_periodic}
\end{equation}
The classical ground states are easily obtained:
\begin{equation}
\Phi=\frac{2\pi}{\beta}n\:,\: n=0,\ldots,k-1\label{classical_vacua}
\end{equation}
which shows that the condition (\ref{period}) corresponds to identifying
the minima of the cosine potential with a period \ensuremath{k}
. In the infinite volume (\ensuremath{L=\infty}
) quantum theory these correspond to vacuum states \ensuremath{\left|n\right\rangle }
 which have the property
\begin{equation}
\left\langle n\right|\Phi(x,t)\left|n\right\rangle =\frac{2\pi}{\beta}n\qquad n=0,1,\dots,k-1\label{n_vacua}
\end{equation}
These states are all degenerate in the classical theory and also at
quantum level when \ensuremath{L=\infty}
; however, tunneling lifts the degeneracy in finite volume \ensuremath{L<\infty}
. The spectrum of breather multi-particle states is obtained in $k$
copies corresponding to the $k$ vacua, and in finite volume tunneling
splits the degeneracy between these states by an amount of order $\mbox{e}^{-ML}$
where $M$ is the soliton mass. In the $k$-folded model, the local
operators of most interest are the exponential fields
\[
\mathrm{e}^{i\frac{n}{k}\beta\Phi}\qquad n\in\mathbb{Z}
\]
In the sequel the value of $k$ is set to $2$ so the model considered
is the $2$-folded sine-Gordon theory \ensuremath{SG(\beta,2)}
.

\section{Form factors in finite volume\label{sec:Form-factors-in}}

\subsection{Corrections to all order in $1/L$\label{sub:Corrections-to-all-in-1overL}}

In this work we restrict our attention to states containing only breathers
(i.e. no solitons). As breathers are singlets and therefore scatter
diagonally, the formulae derived by Pozsgay and Takács in \cite{Pozsgay:2007kn,Pozsgay:2007gx}
are directly applicable. The first ingredient is to describe the multi-breather
energy levels corresponding to the states
\[
\left|B_{r_{1}}(\theta_{1})\dots B_{r_{N}}(\theta_{N})\right\rangle 
\]
whose finite volume counter part we are going to label
\[
\vert\{I_{1},\dots,I_{N}\}\rangle_{r_{1}\dots r_{N},L}
\]
where the $I_{k}$ are the momentum quantum numbers. In a finite volume
$L$, momentum quantization is governed (up to corrections decreasing
exponentially with $L$) by the Bethe-Yang equations:
\begin{equation}
Q_{k}(\theta_{1},\dots,\theta_{n})=m_{r_{k}}L\sinh\theta_{k}+\sum_{j\neq k}\delta_{r_{j}r_{k}}\left(\theta_{k}-\theta_{j}\right)=2\pi I_{k}\qquad I_{k}\in\mathbb{Z}\label{eq:breatherby}
\end{equation}
where the phase-shift is defined as 
\[
S_{rs}(\theta)=\mathrm{e}^{i\delta_{sr}(\theta)}
\]
These equations are nothing else than an appropriate extension of
the description of two-particle scattering states in finite volume
\cite{Luscher:1986pf,Luscher:1990ux}. In general quantum field theories
this description is only valid under the inelastic threshold, but
due to integrability it can be extended to all multi-particle states
regardless of their energy. 

In practical calculations, because breathers of the same species satisfy
an effective exclusion rule due to
\[
S_{rr}(0)=-1
\]
it is best to redefine phase-shifts corresponding to them by extracting
a minus sign:
\[
S_{rr}(\theta)=-\mathrm{e}^{i\delta_{rr}(\theta)}
\]
so that $\delta_{sr}(0)=0$ can be taken for all $s,r$ and all the
phase-shifts can be defined as continuous functions over the whole
real $\theta$ axis. This entails shifting appropriate quantum numbers
$I_{k}$ to half-integer values. Given a solution $\tilde{\theta}_{1},\dots,\tilde{\theta}_{N}$
to the quantization relations (\ref{eq:breatherby}) the energy and
the momentum of the state can be written as 
\begin{eqnarray*}
E & = & \sum_{k=1}^{N}m_{r_{k}}\cosh\tilde{\theta}_{k}\\
P & = & \sum_{k=1}^{N}m_{r_{k}}\sinh\tilde{\theta}_{k}=\frac{2\pi}{L}\sum_{k}I_{k}
\end{eqnarray*}
(using that -- with our choice of the phase-shift functions -- unitarity
entails $\delta_{sr}(\theta)+\delta_{rs}(-\theta)=0$). The rapidity-space
density of $n$-particle states can be calculated as 
\begin{equation}
\rho_{r_{1}\dots r_{n}}(\theta_{1},\dots,\theta_{n})=\det\mathcal{J}^{(n)}\qquad,\qquad\mathcal{J}_{kl}^{(n)}=\frac{\partial Q_{k}(\theta_{1},\dots,\theta_{n})}{\partial\theta_{l}}\quad,\quad k,l=1,\dots,n\label{eq:byjacobian}
\end{equation}
In infinite volume the matrix elements of local operators between
multi-particle states composed of breathers can be expressed in terms
of the elementary form factor functions
\begin{equation}
F_{r_{1}\dots r_{N}}^{\mathcal{O}}(\theta_{1},\dots,\theta_{N})=\langle0|\mathcal{O}|B_{r_{1}}(\theta_{1})\dots B_{r_{N}}(\theta_{N})\rangle\label{eq:elementaryff}
\end{equation}
using crossing symmetry
\begin{align}
\langle B_{s_{1}}(\theta_{1}')\dots B_{s_{M}}(\theta_{M}')|\mathcal{O}|B_{r_{1}}(\theta_{1})\dots B_{r_{N}}(\theta_{N})\rangle & =F_{s_{M}\dots s_{1}r_{1}\dots r_{N}}^{\mathcal{O}}(\theta_{M}'+i\pi,\dots,\theta_{1}'+i\pi,\tilde{\theta}_{1},\dots,\tilde{\theta}_{N})\nonumber \\
 & +\mbox{disconnected contributions}\label{eq:crossing}
\end{align}
where disconnected contributions only arise when there are particles
in the two states that have identical quantum numbers and momenta. 

In terms of the elementary form factors (\ref{eq:elementaryff}),
the finite volume matrix elements of local operators between multi-particle
states can be written as \cite{Pozsgay:2007kn} 
\begin{eqnarray}
 &  & \left|\,_{s_{1}\dots s_{M}}\langle\{I_{1}',\dots,I_{M}'\}\vert\mathcal{O}(0,0)\vert\{I_{1},\dots,I_{N}\}\rangle_{r_{1}\dots r_{N},L}\right|=\nonumber \\
 &  & \qquad\left|\frac{F_{s_{M}\dots s_{1}r_{1}\dots r_{N}}^{\mathcal{O}}(\tilde{\theta}_{M}'+i\pi,\dots,\tilde{\theta}_{1}'+i\pi,\tilde{\theta}_{1},\dots,\tilde{\theta}_{N})}{\sqrt{\rho_{r_{1}\dots r_{N}}(\tilde{\theta}_{1},\dots,\tilde{\theta}_{N})\rho_{s_{1}\dots s_{M}}(\tilde{\theta}_{1}',\dots,\tilde{\theta}_{M}')}}\right|+O(\mathrm{e}^{-\mu L})\label{eq:genffrelation}
\end{eqnarray}
which is valid provided there are no disconnected terms. This is essentially
the extension of the Lellouch-Lüscher formalism \cite{Lellouch:2000pv}
to general multi-particle states. The finite volume quantization relations
in (\ref{eq:breatherby}) and the expression of density factors $\rho$
in (\ref{eq:byjacobian}) are specific to the setting of two-dimensional
integrable field theories, but the relation (\ref{eq:genffrelation})
which states that finite volume matrix elements differ from their
infinite volume counterparts by normalization factors given by the
square roots of densities of states is generally valid in any quantum
field theory (with a nonzero mass gap), as can be seen by considering
the way this expression was derived in \cite{Pozsgay:2007kn}.

Note the absolute values in (\ref{eq:genffrelation}) that are necessary
to take into account that the finite volume and infinite volume phase
conventions differ. In our framework it is known how to compensate
for the difference. The finite volume eigenvectors obtained from TCSA
can be chosen real, as the Hamiltonian is a real symmetric matrix,
and then all matrix elements of exponential operators turn out to
be real. On the other hand, apart from the $i$ factors in (\ref{eq:b1ffs})
the only complex function is the minimal two-particle form factor
$f_{\xi}(\theta)$, whose phase at a real value of the argument is
equal to the square root of the $S$-matrix. Therefore the above relation
can be rewritten as
\begin{eqnarray}
 &  & \,_{s_{1}\dots s_{M}}\langle\{I_{1}',\dots,I_{M}'\}\vert\mathcal{O}(0,0)\vert\{I_{1},\dots,I_{N}\}\rangle_{r_{1}\dots r_{N},L}=\nonumber \\
 &  & \qquad\pm\frac{F_{s_{M}\dots s_{1}r_{1}\dots r_{N}}^{\mathcal{O}}(\tilde{\theta}_{M}'+i\pi,\dots,\tilde{\theta}_{1}'+i\pi,\tilde{\theta}_{1},\dots,\tilde{\theta}_{N})}{\sqrt{\rho_{r_{1}\dots r_{N}}(\tilde{\theta}_{1},\dots,\tilde{\theta}_{N})\rho_{s_{1}\dots s_{M}}(\tilde{\theta}_{1}',\dots,\tilde{\theta}_{M}')}}\times\label{eq:genffrelationphasecorr}\\
 &  & \qquad i^{-(N+M)}\sqrt{\prod_{1\leq k<l\leq M}S_{s_{k}s_{l}}(\tilde{\theta}_{k}'-\tilde{\theta}_{l}')\prod_{1\leq k<l\leq N}S_{r_{k}r_{l}}(\tilde{\theta}_{k}-\tilde{\theta}_{l})}+O(\mathrm{e}^{-\mu L})\nonumber 
\end{eqnarray}
where the only remaining ambiguity is a sign corresponding to the
choice of the branch of the square root (note that this ambiguity
is also manifest in the TCSA eigenvectors since demanding their reality
does not fix their sign). Note also that the values 
\[
f_{\xi}(\theta+i\pi)
\]
are real whenever $\theta$ is real.

Disconnected terms may appear when there are two breathers of the
same species in the two states whose rapidities exactly coincide.
This can happen in two cases: when the two states are identical, or
when there is a particle with exactly zero momentum in both of them
\cite{Pozsgay:2007gx}. Here we only quote the latter since that is
needed for our computations. Defining the function
\begin{eqnarray}
 &  & \mathcal{F}_{k,l}(\theta_{1}',\dots,\theta_{k}'|\theta_{1},\dots,\theta_{l})=\nonumber \\
 &  & \lim_{\epsilon\rightarrow0}F_{\underbrace{{\scriptstyle 11\dots1}}_{2k+2l+2}}(i\pi+\theta_{1}'+\epsilon,\dots,i\pi+\theta_{k}'+\epsilon,i\pi-\theta_{k}'+\epsilon,\dots,i\pi-\theta_{1}'+\epsilon,i\pi+\epsilon,\nonumber \\
 &  & 0,\theta_{1},\dots,\theta_{l},-\theta_{l},\dots,-\theta_{1})\label{eq:oddoddlimitdef}
\end{eqnarray}
one can write
\begin{eqnarray}
 &  & \,_{\underbrace{{\scriptstyle 11\dots1}}_{2k+1}}\langle\{I_{1}',\dots,I_{k}',0,-I_{k}',\dots,-I_{1}'\}|\Phi|\{I_{1},\dots,I_{l},0,-I_{l},\dots,-I_{1}\}\rangle_{\underbrace{{\scriptstyle 11\dots1}}_{2l+1},L}\label{eq:oddoddrule}\\
 & = & \frac{\Big(\mathcal{F}_{k,l}(\tilde{\theta}_{1}',\dots,\tilde{\theta}_{k}'|\tilde{\theta}_{1},\dots,\tilde{\theta}_{l})+mL\, F_{\underbrace{{\scriptstyle 11\dots1}}_{2k+2l}}(i\pi+\tilde{\theta}_{1}',\dots,i\pi-\tilde{\theta}_{1}',\tilde{\theta}_{1},\dots,-\tilde{\theta}_{1})\Big)}{\sqrt{\rho_{2k+1}(\tilde{\theta}_{1}',\dots,\tilde{\theta}_{k}',0,-\tilde{\theta}_{k}',\dots,-\tilde{\theta}_{1}')\rho_{2l+1}(\tilde{\theta}_{1},\dots,\tilde{\theta}_{l},0,-\tilde{\theta}_{l},\dots,-\tilde{\theta}_{1})}}\times\nonumber \\
 &  & (\mbox{phase factor})+O(\mathrm{e}^{-\mu L})\nonumber 
\end{eqnarray}
where the compensating phase factor is entirely analogous to the one
in (\ref{eq:genffrelationphasecorr}). 

The above predictions for finite volume energy levels and matrix elements
are expected to be exact to all (finite) orders in $1/L$ \cite{Pozsgay:2007kn,Pozsgay:2007gx}
(note that the exponential corrections are non-analytic in this variable).

\subsection{\label{sub:mu-terms-the-leading-exp-correction}$\mu$-terms: the
leading exponential corrections}

\subsubsection{$\mu$-terms for energy levels\label{sub:mu-terms-for-b2}}

Using the ideas in \cite{Pozsgay:2008bf}, we can model a finite volume
$B_{2}$ state as a pair of $B_{1}$ particles with complex conjugate
rapidities
\[
B_{2}(\theta)\sim B_{1}(\theta+iu)B_{1}(\theta-iu)
\]
where $u$ can be obtained by solving the $B_{1}B_{1}$ Bethe-Yang
equations (written here in exponential form):
\begin{align}
\mathrm{e}^{im_{1}L\sinh(\theta\pm iu)}S_{11}(\pm2iu) & =1\label{eq:b2_b1b1_by}
\end{align}
where 
\[
S_{11}(\theta)=\frac{\sinh\theta+i\sin\pi\xi}{\sinh\theta-i\sin\pi\xi}
\]
The solution for $u$ has the large volume behavior 
\begin{equation}
u\sim\frac{\pi\xi}{2}+\tan\frac{\pi\xi}{2}\mathrm{e}^{-\mu_{11}^{2}L\cosh\theta}\label{eq:u_asymptotics}
\end{equation}
with 
\[
\mu_{11}^{2}=\sqrt{m_{1}^{2}-\frac{m_{2}^{2}}{4}}=m_{1}\sin\frac{\pi\xi}{2}
\]
Therefore, for $L\rightarrow\infty$ one obtains the usual bootstrap
identification
\[
B_{2}(\theta)\simeq B_{1}(\theta+i\pi\xi/2)B_{1}(\theta-i\pi\xi/2)
\]
and the momentum and energy of the state tends to
\begin{align*}
m_{1}\sinh(\theta+iu)+m_{1}\sinh(\theta-iu) & =2m_{1}\cos u\sinh\theta\mathop{\rightarrow}_{L\rightarrow\infty}m_{2}\sinh\theta\\
m_{1}\cosh(\theta+iu)+m_{1}\cosh(\theta-iu) & =2m_{1}\cos u\cosh\theta\mathop{\rightarrow}_{L\rightarrow\infty}m_{2}\cosh\theta
\end{align*}
where 
\[
m_{2}=2m_{1}\cos\frac{\pi\xi}{2}
\]
is the mass of \textbf{$B_{2}$} in terms of that of $B_{1}$, consistent
with (\ref{eq:breather_mass}). For a stationary $B_{2}$, the energy
dependence on the volume is 
\begin{equation}
E_{2}(L)=2m_{1}\cos u\sim m_{2}-\left(\gamma_{11}^{2}\right)^{2}\mu_{11}^{2}\mathrm{e}^{-\mu_{11}^{2}L}+\dots\label{eq:b2mupred}
\end{equation}
where the second expression shows the asymptotic behavior valid for
large enough $L$. One can also write
\[
\mu_{11}^{2}=m_{1}\sin u_{12}^{1}
\]
where the $B_{1}B_{2}\rightarrow B_{1}$ fusion angle is
\[
u_{12}^{1}=\pi\left(1-\frac{\xi}{2}\right)
\]
\begin{figure}
\begin{centering}
\begin{tabular}{ccc}
\includegraphics[scale=0.8]{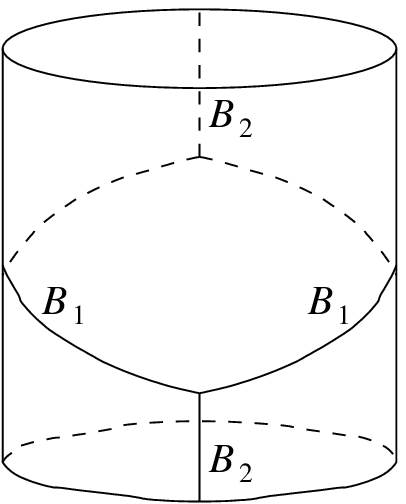} & ~~~~~~~~~~~~~~~~~~~~~ & \includegraphics[scale=0.8]{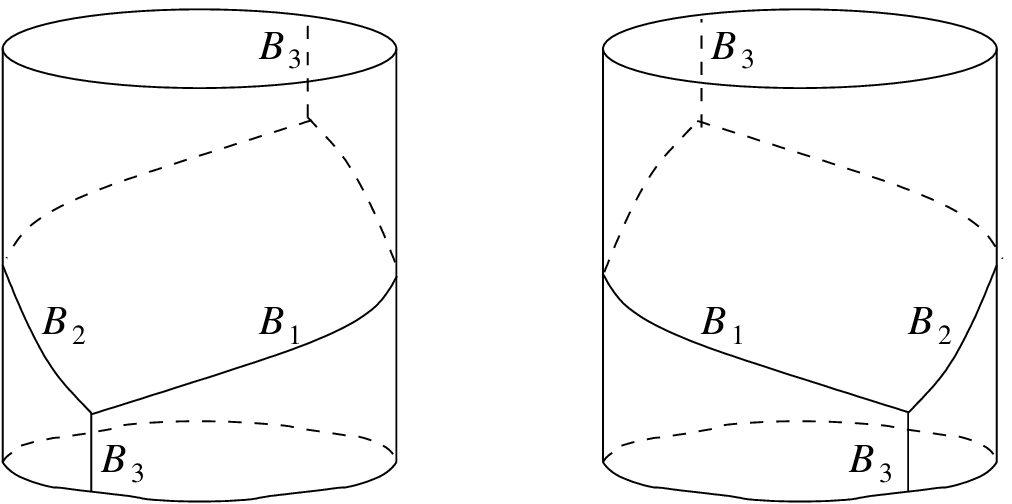}\tabularnewline
(a) $B_{2}$ &  & (b) $B_{3}$\tabularnewline
\end{tabular}
\par\end{centering}

\caption{\label{fig:b2b3muterm}$\mu$-term diagrams}
\end{figure}
The asymptotic behaviour in (\ref{eq:b2mupred}) coincides with that
predicted by Lüscher's finite volume mass formula \cite{Luscher:1985dn},
which was worked out in more detail for $2$-dimensional quantum field
theory in \cite{Klassen:1990ub} and is illustrated graphically in
Fig. \ref{fig:b2b3muterm} (a). Here $\gamma_{11}^{2}$ is the three-particle
coupling defined in (\ref{eq:gammacoeffs}). The kinematical variables
can be represented by the mass triangle shown in Fig. \ref{fig:Mass-triangle,-fusion},
with $\mu_{ab}^{c}$ being the height of the triangle perpendicular
to the side $m_{c}$. Note that $\mu_{ab}^{c}$ can be much smaller
than other mass scales in the theory if the binding energy is low,
i.e. when $m_{c}\sim m_{a}+m_{b}$.

\begin{figure}
\begin{centering}
\includegraphics{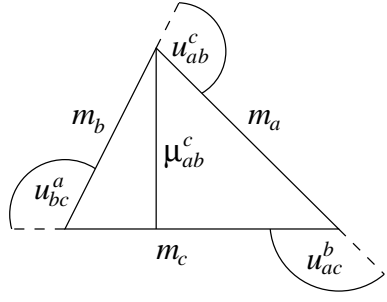}
\par\end{centering}

\caption{\label{fig:Mass-triangle,-fusion} Mass triangle, fusion angles and
$\mu$ parameter}
\end{figure}

A stationary third breather $B_{3}$ can be modeled as a composite
of three $B_{1}$ particles
\[
B_{3}(\theta=0)\sim B_{1}(iu)B_{1}(0)B_{1}(-iu)
\]
where the (nontrivial) Bethe-Yang equations read 
\begin{align}
\mathrm{e}^{im_{1}L\sinh(\pm iu)}S_{11}(\pm iu)S_{11}(\pm2iu) & =1\label{eq:B3mutermby}
\end{align}
leading to the asymptotic expression
\begin{equation}
u\sim\pi\xi+\frac{4\sin\pi\xi+2\tan\pi\xi}{2\cos\pi\xi-1}\mathrm{e}^{-m_{1}L\sin\pi\xi}\label{eq:b3u_asymptotics}
\end{equation}
The energy of the $B_{3}$ level then becomes
\begin{equation}
E_{3}(L)=m_{1}(1+2\cos u)\sim m_{3}-2\left(\gamma_{12}^{3}\right)^{2}\mu_{12}^{3}\mathrm{e}^{-\mu_{12}^{3}L}+\dots\label{b3mupred}
\end{equation}
where the three-particle coupling $\gamma_{12}^{3}$ is defined in
(\ref{eq:gammacoeffs}) and 
\[
\mu_{12}^{3}=m_{1}\sin u_{13}^{2}=m_{2}\sin u_{23}^{1}
\]
with the fusion angles 
\[
u_{13}^{2}=\pi(1-\xi)\qquad u_{23}^{1}=\pi\left(1-\frac{\xi}{2}\right)
\]
The asymptotic behavior in (\ref{b3mupred}) is again consistent with
the expression given by Lüscher \cite{Luscher:1985dn,Klassen:1990ub},
the factor of $2$ is due to the presence of two processes contributing
the same amount as shown in Fig. \ref{fig:b2b3muterm} (b).

\subsubsection{$\mu$-term corrections for the form factors }

Using the results of \cite{Pozsgay:2008bf}, these can be computed
by continuing (\ref{eq:genffrelationphasecorr}) to complex rapidities
that correspond to representing $B_{2}$ and $B_{3}$ as composites
made out of $B_{1}$. One obtains for vacuum-$B_{2}$ matrix elements
the form
\begin{equation}
\langle0\vert\mathcal{O}(0,0)\vert\{I\}\rangle_{2,L}=\pm\frac{\sqrt{S_{11}(2i\tilde{u})}F_{11}^{\mathcal{O}}(\tilde{\theta}-i\tilde{u},\tilde{\theta}+i\tilde{u})}{\sqrt{\rho_{11}(\tilde{\theta}+i\tilde{u},\tilde{\theta}-i\tilde{u})}}+\dots\label{eq:vacb2ffmu}
\end{equation}
where $\tilde{\theta}$, $\tilde{u}$ is the solution of (\ref{eq:b2_b1b1_by})
with the correct momentum, i.e. 
\[
m_{1}L\sinh(\tilde{\theta}+i\tilde{u})+m_{1}L\sinh(\tilde{\theta}-i\tilde{u})=2\pi I
\]
and the dots denote further (and generally much smaller) exponential
corrections; the $\pm$ sign accounts for the remaining phase ambiguity
from the square root. One can similarly evaluate $B_{1}-B_{2}$ and
$B_{1}B_{1}-B_{2}$ matrix elements using
\begin{eqnarray}
_{1}\langle\{I'\}\vert\mathcal{O}(0,0)\vert\{I\}\rangle_{2,L} & = & \pm i\frac{\sqrt{S_{11}(2i\tilde{u})}F_{111}^{\mathcal{O}}(i\pi+\tilde{\theta}',\tilde{\theta}-i\tilde{u},\tilde{\theta}+i\tilde{u})}{\sqrt{m_{1}L\cosh\tilde{\theta}'\,\rho_{11}(\tilde{\theta}+i\tilde{u},\tilde{\theta}-i\tilde{u})}}+\dots\label{eq:b1b2ffmu}\\
_{11}\langle\{I_{1}',I_{2}'\}\vert\mathcal{O}(0,0)\vert\{I\}\rangle_{2,L} & = & \pm\Bigg[\frac{\sqrt{S_{11}(2i\tilde{u})S_{11}(\tilde{\theta}_{1}'-\tilde{\theta}_{2}')}}{\sqrt{\rho_{11}(\tilde{\theta}_{1}',\tilde{\theta}_{2}')\,\rho_{11}(\tilde{\theta}+i\tilde{u},\tilde{\theta}-i\tilde{u})}}\times\nonumber \\
 &  & F_{1111}^{\mathcal{O}}(i\pi+\tilde{\theta}_{2}',i\pi+\tilde{\theta}_{1}',\tilde{\theta}-i\tilde{u},\tilde{\theta}+i\tilde{u})\Bigg]+\dots\nonumber 
\end{eqnarray}
For large enough $L$, we can use (\ref{eq:u_asymptotics}) and keep
the leading term of these expressions in the limit $\tilde{u}\rightarrow\pi\xi/2$
to obtain
\begin{eqnarray}
\langle0\vert\mathcal{O}(0,0)\vert\{I\}\rangle_{2,L} & = & \pm\frac{F_{2}^{\mathcal{O}}(\tilde{\theta})}{\sqrt{m_{2}L\cosh\tilde{\theta}}}+O\left(\mathrm{e}^{-\mu_{11}^{2}L}\right)\label{eq:b2ffwithoutmu}\\
_{1}\langle\{I'\}\vert\mathcal{O}(0,0)\vert\{I\}\rangle_{2,L} & = & \pm i\frac{F_{12}^{\mathcal{O}}(i\pi+\tilde{\theta}',\tilde{\theta})}{\sqrt{m_{1}L\cosh\tilde{\theta}'\, m_{2}L\cosh\tilde{\theta}}}+O\left(\mathrm{e}^{-\mu_{11}^{2}L}\right)\nonumber \\
_{11}\langle\{I_{1}',I_{2}'\}\vert\mathcal{O}(0,0)\vert\{I\}\rangle_{2,L} & = & \pm\frac{\sqrt{S_{11}(\tilde{\theta}_{1}'-\tilde{\theta}_{2}')}F_{12}^{\mathcal{O}}(i\pi+\tilde{\theta}_{2}',i\pi+\tilde{\theta}_{1}',\tilde{\theta})}{\sqrt{\rho_{11}(\tilde{\theta}_{1}'-\tilde{\theta}_{2}')\, m_{2}L\cosh\tilde{\theta}}}\times\nonumber \\
 &  & +O\left(\mathrm{e}^{-\mu_{11}^{2}L}\right)\nonumber 
\end{eqnarray}
where we also used (\ref{eq:higherbff}). This is just what we obtain
from the general expression (\ref{eq:genffrelationphasecorr}). In
performing this calculation, the singular behavior of the factor $S_{11}(2i\tilde{u})$
for $\tilde{u}\rightarrow\pi\xi/2$ is canceled by a singular term
from the denominator factor 
\[
\rho_{11}(\tilde{\theta}+i\tilde{u},\tilde{\theta}-i\tilde{u})=\frac{m_{2}L\cosh\tilde{\theta}}{2\left(\tilde{u}-\frac{\pi\xi}{2}\right)}+\mbox{regular terms}
\]
For matrix elements involving a stationary $B_{3}$ we obtain
\begin{eqnarray}
\langle0\vert\mathcal{O}(0,0)\vert\{0\}\rangle_{3,L} & = & \pm i\frac{\sqrt{S_{11}(2i\tilde{u})}S_{11}(i\tilde{u})F_{111}^{\mathcal{O}}(-i\tilde{u},0,+i\tilde{u})}{\sqrt{\rho_{111}(+i\tilde{u},0,-i\tilde{u})}}+\dots\label{eq:b3ffmu}\\
_{1}\langle\{I'\}\vert\mathcal{O}(0,0)\vert\{0\}\rangle_{3,L} & = & \pm\frac{\sqrt{S_{11}(2i\tilde{u})}S_{11}(i\tilde{u})F_{1111}^{\mathcal{O}}(i\pi+\tilde{\theta}',-i\tilde{u},0,+i\tilde{u})}{\sqrt{m_{1}L\cosh\tilde{\theta}'\,\rho_{111}(+i\tilde{u},0,-i\tilde{u})}}+\dots\nonumber \\
_{11}\langle\{I_{1}',I_{2}'\}\vert\mathcal{O}(0,0)\vert\{0\}\rangle_{3,L} & = & \pm i\Bigg[\frac{\sqrt{S_{11}(2i\tilde{u})S_{11}(\tilde{\theta}_{1}'-\tilde{\theta}_{2}')}S_{11}(i\tilde{u})}{\sqrt{\rho_{11}(\tilde{\theta}_{1}',\tilde{\theta}_{2}')\,\rho_{111}(+i\tilde{u},0,-i\tilde{u})}}\times\nonumber \\
 &  & F_{11111}^{\mathcal{O}}(i\pi+\tilde{\theta}_{2}',i\pi+\tilde{\theta}_{1}',-i\tilde{u},0,+i\tilde{u})\Bigg]+\dots\nonumber 
\end{eqnarray}
where $\tilde{u}$ solves (\ref{eq:B3mutermby}) with the asymptotics
(\ref{eq:b3u_asymptotics}). There is an exception to the second formula,
however: when $I'=0$, the $B_{1}$ is stationary and there is a disconnected
piece, which according to (\ref{eq:oddoddrule}) leads to the modification
\begin{equation}
_{1}\langle\{0\}\vert\mathcal{O}(0,0)\vert\{0\}\rangle_{3,L}=\pm\frac{\sqrt{S_{11}(2i\tilde{u})}S_{11}(i\tilde{u})\bar{\mathcal{F}}_{1,3}^{\mathcal{O}}(-i\tilde{u})+m_{1}L\,\sqrt{S_{11}(2i\tilde{u})}F_{11}^{\mathcal{O}}(-i\tilde{u},+i\tilde{u})}{\sqrt{m_{1}L\cosh\tilde{\theta}'\,\rho_{111}(+i\tilde{u},0,-i\tilde{u})}}+\dots\label{eq:b1b3ffimprmu}
\end{equation}
where 
\[
\bar{\mathcal{F}}_{1,3}^{\mathcal{O}}(\tilde{\theta})=\lim_{\epsilon\rightarrow0}F_{1111}^{\mathcal{O}}(i\pi+\epsilon,\tilde{\theta},0,-\tilde{\theta})
\]
Using the exchange property
\begin{eqnarray}
 &  & F_{i_{1}\dots i_{k}i_{k+1}\dots i_{n}}^{\mathcal{O}}(\theta_{1},\dots,\theta_{k},\theta_{k+1},\dots,\theta_{n})=\nonumber \\
 &  & \qquad S_{i_{k}i_{k+1}}(\theta_{k}-\theta_{k+1})F_{i_{1}\dots i_{k+1}i_{k}\dots i_{n}}^{\mathcal{O}}(\theta_{1},\dots,\theta_{k+1},\theta_{k},\dots,\theta_{n})\label{eq:exchangeaxiom}
\end{eqnarray}
we can write 
\begin{align*}
\bar{\mathcal{F}}_{1,3}^{\mathcal{O}}(\tilde{\theta}) & =S(-\tilde{\theta})\mathcal{F}_{1,3}^{\mathcal{O}}(\tilde{\theta})\\
\mbox{where \,} & \mathcal{F}_{1,3}^{\mathcal{O}}(\tilde{\theta})=\lim_{\epsilon\rightarrow0}F_{1111}^{\mathcal{O}}(i\pi+\epsilon,0,\tilde{\theta},-\tilde{\theta})
\end{align*}
which explains the presence of the factor $S_{11}(i\tilde{u})$ in
the coefficient of $\bar{\mathcal{F}}_{1,3}^{\mathcal{O}}$ in (\ref{eq:b1b3ffimprmu}),
in contrast to formula (\ref{eq:oddoddrule}). (The factor $\sqrt{S_{11}(2i\tilde{u})}$,
common to both terms results from the analytic continuation of the
compensating phase factors discussed in the previous subsection, as
all similar factors in other cases do as well.)

It can again be verified that for $L$ large enough, taking the limit
$\tilde{u}\rightarrow\pi\xi$ again results in agreement with what
is expected from (\ref{eq:genffrelationphasecorr}):
\begin{eqnarray}
\langle0\vert\mathcal{O}(0,0)\vert\{0\}\rangle_{3,L} & = & \pm\frac{F_{3}^{\mathcal{O}}(0)}{\sqrt{m_{3}L}}+O\left(\mathrm{e}^{-\mu_{12}^{3}L}\right)\label{eq:b3ffwithoutmu}\\
_{1}\langle\{I'\}\vert\mathcal{O}(0,0)\vert\{0\}\rangle_{3,L} & = & \pm i\frac{F_{13}^{\mathcal{O}}(i\pi+\tilde{\theta}',0)}{\sqrt{m_{1}L\cosh\tilde{\theta}'\, m_{3}L}}+O\left(\mathrm{e}^{-\mu_{12}^{3}L}\right)\nonumber \\
_{11}\langle\{I_{1}',I_{2}'\}\vert\mathcal{O}(0,0)\vert\{0\}\rangle_{3,L} & = & \pm\frac{\sqrt{S_{11}(\tilde{\theta}_{1}'-\tilde{\theta}_{2}')}F_{13}^{\mathcal{O}}(i\pi+\tilde{\theta}_{2}',i\pi+\tilde{\theta}_{1}',0)}{\sqrt{\rho_{11}(\tilde{\theta}_{1}'-\tilde{\theta}_{2}')\, m_{3}L}}\times\nonumber \\
 &  & +O\left(\mathrm{e}^{-\mu_{12}^{3}L}\right)\nonumber 
\end{eqnarray}
Note that the additional term in (\ref{eq:b1b3ffimprmu}) drops out
because the factor $S_{11}(2i\tilde{u})$ has no pole to compensate
for the singular behavior of the denominator which is of the form
\[
\rho_{111}(+i\tilde{u},0,-i\tilde{u})\sim\frac{m_{3}L}{(\tilde{u}-\pi\xi)^{2}}+\mbox{less singular terms}
\]

\subsection{Beyond the leading $\mu$-terms}

It is important to consider subleading exponential corrections. First
of all, the breathers also have $\mu$-terms corresponding to their
description as soliton-antisoliton bound states, with the characteristic
scale 
\[
\mu_{s\bar{s}}^{k}=\sqrt{M^{2}-\frac{m_{k}^{2}}{4}}
\]
In addition to the $\mu$-terms there are corrections which correspond
to particle loops winding around the space-time cylinder. They are
called $F$-terms because they depend on the forward scattering amplitude
$F$ continued to rapidity difference with imaginary part $i\pi/2$
\cite{Luscher:1985dn} and are suppressed by the mass of the particle
propagating in the loop. Last, but not least, there are also the tunneling
corrections related to the presence of multiple vacua ($k=2$ in the
case considered below), suppressed by the soliton mass $M$, which
are essentially vacuum versions of $F$-terms. These contributions
were evaluated using the dilute instanton gas approximation in \cite{Bajnok:2000wm}.
$F$-terms contributions are illustrated in Fig. and their scales
are compared explicitly to $\mu$-term scales at the beginning of
the next section.

\begin{figure}
\begin{centering}
\begin{tabular}{ccc}
\includegraphics[scale=0.8]{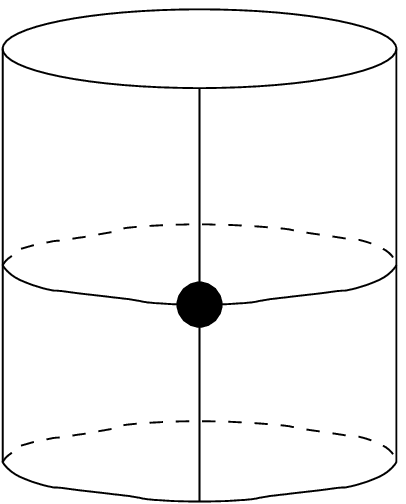} & ~~~~~~~~~~~~~~~~~~~~~ & \includegraphics[scale=0.8]{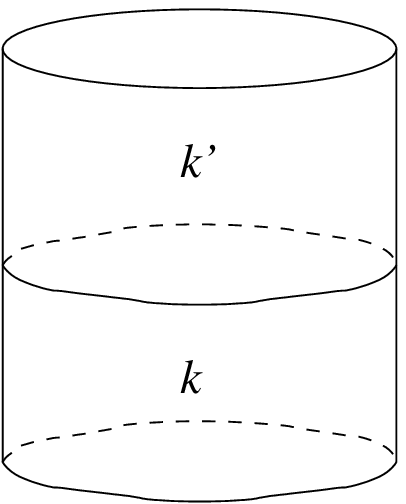}\tabularnewline
(a)  &  & (b) \tabularnewline
\end{tabular}
\par\end{centering}

\caption{\label{fig:Fterms}(a) $F$-term correction to the mass, the filled
circle denotes the forward scattering amplitude evaluated at $\theta+i\pi/2$,
where $\theta$ is the rapidity of the particle propagating in the
loop.\protect \\
(b) Vacuum tunneling between two adjacent vacua ($k'=k\pm1$), the
loop is the world line of a soliton propagating around the cylinder.}
\end{figure}

\section{Determining matrix elements of local operators: the effect of $\mu$-terms\label{sec:Determining-matrix-elements}}

To evaluate the form factors numerically, we use the truncated conformal
space approach (TCSA) pioneered by Yurov and Zamolodchikov \cite{Yurov:1989yu}.
The essential idea is to truncate the Hilbert space of the massless
free boson introducing an upper energy cutoff (called the truncation),
which gives a finite dimensional space in finite volume $L$. The
Hamiltonian (\ref{eq:pcft_action}) can then be represented as a finite
numerical matrix and the energy levels evaluated by numerical diagonalization.
Truncation introduces a source of systematic errors, known as truncation
errors, which are analogous to discretization errors in lattice field
theory. Since the wave-functions of the the interacting theory are
expanded in terms of lowest-lying states of the limiting conformal
model (obtained as the high energy limit $L\rightarrow0$), truncation
errors grow with volume and they are also larger for higher lying
levels. In most of the calculations in this paper they are too small
to be considered, with the exception of the issue of resonances treated
in section \ref{sec:Decay-widths-of}.

The extension of TCSA to the sine-Gordon model was developed in \cite{Feverati:1998va}
and has found numerous applications since then. The Hilbert space
can be split by the eigenvalues of the topological charge $\mathcal{Q}$
(or winding number) and the spatial momentum $P$, where the eigenvalues
of the latter are of the form
\[
\frac{2\pi s}{L}\qquad s\in\mathbb{Z}
\]
In sectors with vanishing topological charge, we can make use of the
symmetry of the Hamiltonian under 
\[
\mathcal{C}:\qquad\Phi(x,t)\rightarrow-\Phi(x,t)
\]
which is equivalent to conjugation of the topological charge. The
truncated space can be split into $\mathcal{C}$-even and $\mathcal{C}$-odd
subspaces that have roughly equal dimensions, which speeds up the
diagonalization of the Hamiltonian. 

Using relation (\ref{eq:mass_scale}) we can express all energy levels
and matrix elements in units of (appropriate powers of) the soliton
mass $M$, and we also introduce the dimensionless volume variable
$l=ML$. The general procedure is the same as in \cite{Pozsgay:2007kn,Pozsgay:2007gx}:
the particle content of energy levels can be identified by matching
the numerical TCSA spectrum against the predictions of the Bethe-Yang
equations (\ref{eq:breatherby}). After identification, one can compare
the appropriate matrix elements to the theoretical values given by
(\ref{eq:genffrelation}).

A detailed study of breather form factors based on the formalism explained
in subsection \ref{sub:Corrections-to-all-in-1overL} has been already
performed in \cite{Feher:2011aa}, therefore it is not repeated here.
For the sake of concentrating on the essence of the matter, the effect
of $\mu$-terms is demonstrated on selected examples. In the process
of the computation, the analysis was performed for a large number
of different matrix elements and also for several values of coupling
constant $\xi$, so the data shown are just a small, but representative
subset of results. In view of the next section (dealing with resonances)
the most interesting operator to study is 
\[
\mathcal{O}=\mathrm{e}^{i\frac{\beta}{2}\Phi(0)}
\]
to which therefore all the form factor plots correspond. The numerical
data presented below correspond to the coupling $\xi=50/311$, for
which it is instructive to compare the scales involved. The masses
of the first three breathers in terms of the soliton mass $M$ are
\begin{align*}
m_{1} & =0.49973\dots\times M\\
m_{2} & =0.96775\dots\times M\\
m_{3} & =1.37439\dots\times M
\end{align*}
The $\mu$-term scales are
\begin{align*}
\mu_{11}^{2} & =0.12486\dots\times M\\
\mu_{12}^{3} & =0.24181\dots\times M\\
\mu_{s\bar{s}}^{1} & =0.96828\dots\times M\\
\mu_{s\bar{s}}^{2} & =0.87514\dots\times M\\
\mu_{s\bar{s}}^{3} & =0.72647\dots\times M
\end{align*}
Therefore it can be seen that the leading exponential corrections
are those driven by the splittings $B_{2}\rightarrow B_{1}B_{1}$
and $B_{3}\rightarrow B_{1}B_{2}$, which are exactly the ones described
by the considerations in subsection \ref{sub:mu-terms-the-leading-exp-correction}.
It is also meaningful to beyond the leading asymptotics to evaluate
them (i.e. to consider the full solution of the associated Bethe-Yang
equations), because even $4\mu_{11}^{2}$ and $2\mu_{12}^{3}$ are
smaller than the next scale $m_{1}$ (although the latter only slightly
for this particular value of the coupling).

\subsection{Finite volume mass corrections \label{sub:Finite-volume-mass}}

\begin{figure}
\begin{centering}
\begin{tabular}{cc}
\includegraphics[scale=0.85]{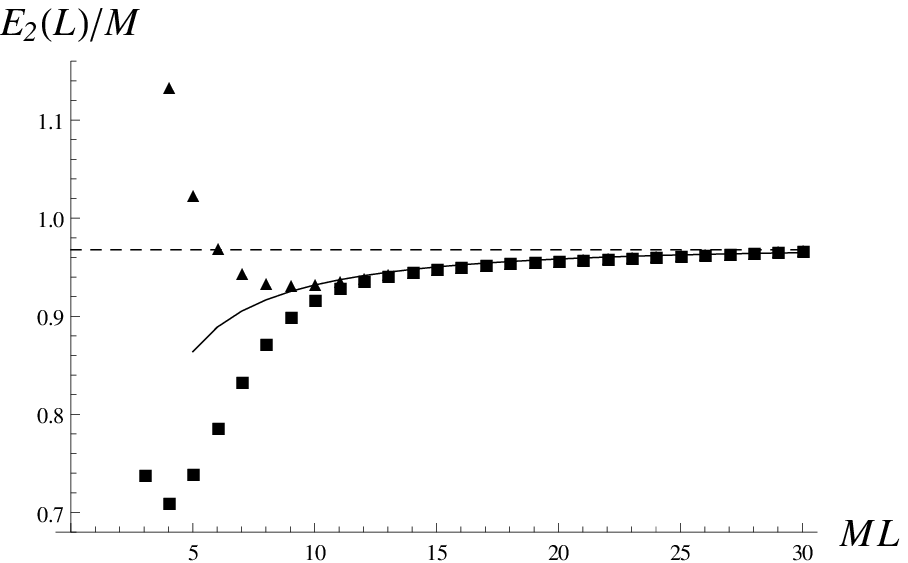} & \includegraphics[scale=0.85]{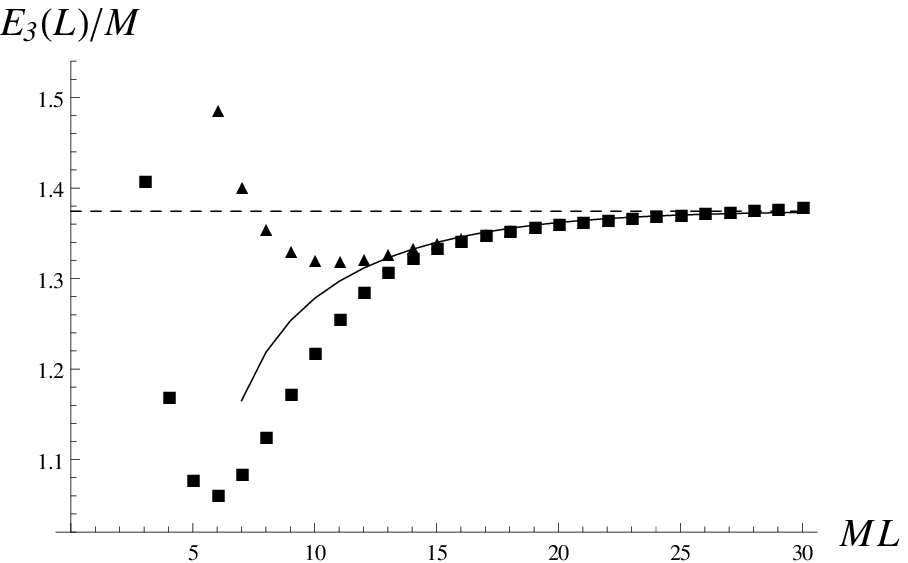}\tabularnewline
(a) $B_{2}$ & (b) $B_{3}$\tabularnewline
\end{tabular}
\par\end{centering}

\caption{\label{fig:massmuterms}$\mu$-term corrections for finite volume
mass at $\xi=50/311$}

\end{figure}

One can start by checking whether the mass corrections predicted by
eqns. (\ref{eq:b2mupred}) for $B_{2}$ and (\ref{b3mupred}) for
$B_{3}$ states agree with the numerical energy levels extracted from
TCSA. Note that there are two copies of zero-momentum $B_{2}$ and
$B_{3}$ levels in a $2$-folded model, so the relevant question is
whether the finite volume mass gaps evaluated from both of them agrees
with a single $\mu$-term correction, at least for volumes that are
large enough for higher exponential terms to be negligible. Indeed,
Fig. \ref{fig:massmuterms} demonstrates that it is so. The dashed
line shows the infinite volume mass value, while the continuous one
is the $\mu$-term corrected behavior. The discrete dots show the
two mass gaps measured in TCSA. 

Note that the correction is rather small (of the order of at most
a few percent) for $ML>10$. Since the smallest particle mass for
the coupling considered is 
\[
m_{1}=2M\sin\frac{\pi\xi}{2}\approx0.5M
\]
this corresponds to volumes larger than five times the correlation
length of the model. According to the considerations in \cite{Lellouch:2000pv},
this should be enough to neglect these corrections in practical computations
(in QCD the role of $m_{1}$ is played by the pion mass $m_{\pi}$).

\subsection{Vacuum--one-particle matrix elements \label{sub:Vacuum--one-particle-matrix-elem}}

\begin{figure}
\begin{centering}
\begin{tabular}{cc}
\includegraphics[scale=0.85]{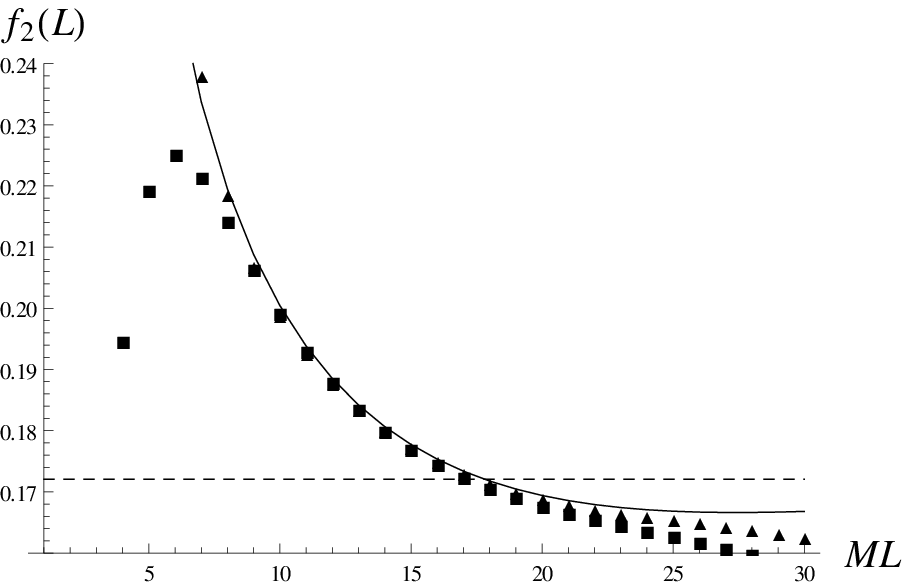} & \includegraphics[scale=0.85]{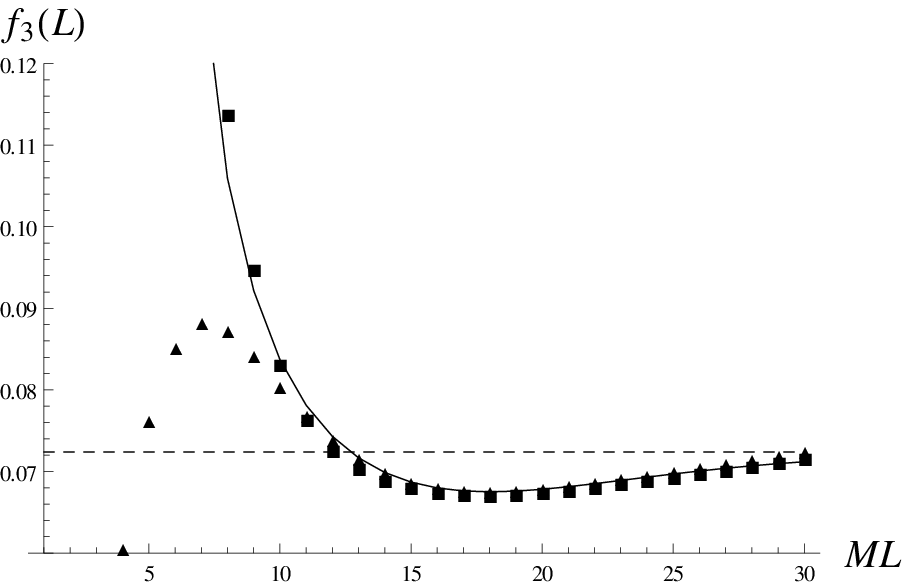}\tabularnewline
(a) $B_{2}$ & (b) $B_{3}$\tabularnewline
\end{tabular}
\par\end{centering}

\caption{\label{fig:vaconepart}$\mu$-term corrections to the vacuum--one-particle
matrix element at $\xi=50/311$. In each case, the dashed line shows
the constant behavior expected when neglecting exponential corrections,
the continuous line is the one including the $\mu$-term corrections,
while the TCSA matrix elements are represented by the two sets of
discrete data points.}
\end{figure}

For these matrix elements, we plot the dimensionless quantity
\begin{equation}
f_{k}(L)=M^{-x}\sqrt{m_{k}L}\langle0|\mathcal{O}|\{0\}\rangle_{k,L}\label{eq:denscorrvac1}
\end{equation}
where 
\[
x=\frac{\beta^{2}}{16\pi}=\frac{\xi}{2(1+\xi)}
\]
is the exact scaling dimension of the operator $\mathcal{O}$. Using
formula (\ref{eq:genffrelation}) this is naively expected to be constant
\begin{equation}
f_{k}(L)=M^{-x}F_{k}^{\mathcal{O}}(0)+O\left(\mathrm{e}^{-\mu L}\right)\label{eq:fkas}
\end{equation}
where 
\[
F_{k}^{\mathcal{O}}(\theta)=\langle0|\mathcal{O}|B_{k}(\theta)\rangle
\]
is the appropriate infinite volume form factor (which is eventually
independent of $\theta$ due to Lorentz invariance). As demonstrated
by Fig. \ref{fig:vaconepart}, however, the exponential corrections
predicted by (\ref{eq:vacb2ffmu}) and the first equation in (\ref{eq:b3ffmu}),
coming from the bound state structure of $B_{2}$ and $B_{3}$ lead
to observable deviations from this behavior, which agrees quite well
with the numerical results. Still, one could say that in the regime
$ML>10$ these are expected to be less than $10\%$, and their presence
can be detected from the fact that $f_{k}(L)$ is not constant. Note
that for each $k$, there are two one-particle $B_{k}$ states, and
with the two vacua this gives four possible matrix elements. However,
two of these are very small due to suppression by tunneling.

\subsection{One-particle--one-particle matrix elements \label{sub:One-particle--one-particle-matri}}

\begin{figure}
\begin{centering}
\begin{tabular}{cc}
\includegraphics[scale=0.85]{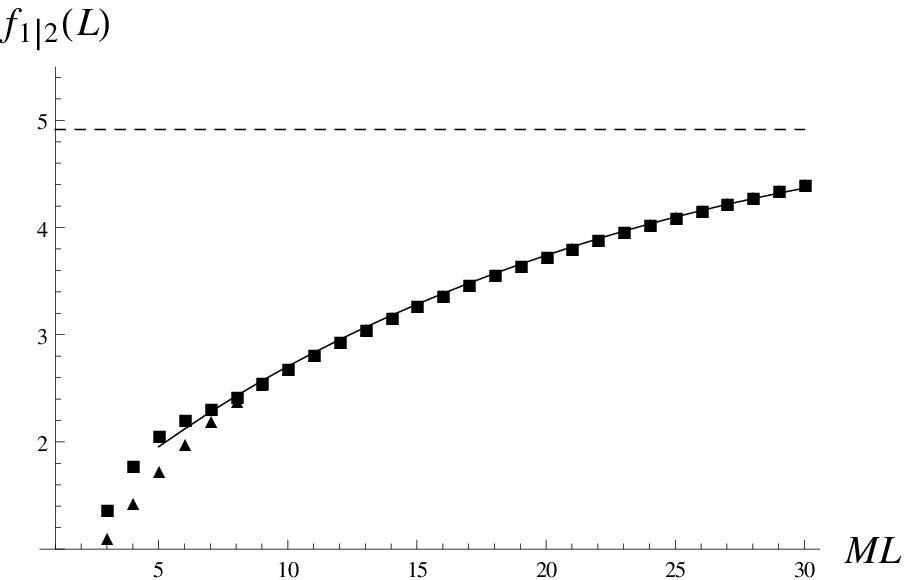} & \includegraphics[scale=0.85]{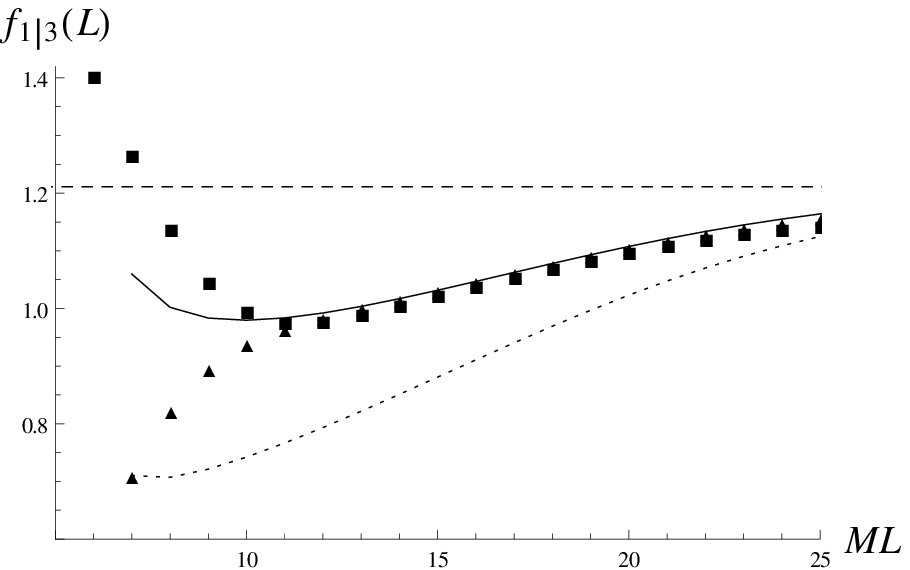}\tabularnewline
(a) $B_{2}$ & (b) $B_{3}$\tabularnewline
\end{tabular}
\par\end{centering}

\caption{\label{fig:onepartonepart}$\mu$-term corrections to the one-particle--one-particle
matrix element at $\xi=50/311$}
\end{figure}

Turning now to matrix elements of the form
\begin{equation}
f_{1|k}(L)=M^{-x}\sqrt{m_{1}L}\sqrt{m_{k}L}\,\,_{1}\langle\{0\}|\mathcal{O}|\{0\}\rangle_{k,L}\label{eq:denscorr11}
\end{equation}
we encounter a surprise. According to (\ref{eq:genffrelation}), such
matrix elements are expected to equal
\begin{equation}
f_{1|k}(L)=M^{-x}F_{1k}^{\mathcal{O}}(i\pi,0)+O\left(\mathrm{e}^{-\mu L}\right)\label{eq:f1kas}
\end{equation}
where 
\[
F_{1k}^{\mathcal{O}}(\theta_{1},\theta_{2})=\langle0|\mathcal{O}|B_{1}(\theta_{1})B_{k}(\theta_{2})\rangle
\]
which under the crossing property (\ref{eq:crossing}) entails that
\[
f_{1|k}(L)=M^{-x}\langle B_{1}(0)|\mathcal{O}|B_{k}(0)\rangle+O\left(\mathrm{e}^{-\mu L}\right)
\]
so this function must be approximately equal to a constant. However,
it is very obvious from Fig. \ref{fig:onepartonepart} that the exponential
corrections are huge (of the order of 100\% for $k=2$) and the constant
value (showed by the dashed line) is approached very slowly. It is
also apparent that the $\mu$-term corrected predictions (the continuous
lines) from (\ref{eq:b1b2ffmu}) and (\ref{eq:b1b3ffimprmu}) describe
the numerical data very well. For the case of $B_{3}$ the ``naive''
$\mu$-term prediction from the second equation in (\ref{eq:b3ffmu})
is also shown with dotted line, demonstrating that it is very important
to get the disconnected parts of the $\mu$-terms right to achieve
agreement.

\subsection{Two-particle--one-particle matrix elements\label{sub:Two-particle--one-particle-matri}}

\begin{figure}
\begin{centering}
\begin{tabular}{cc}
\includegraphics[scale=0.85]{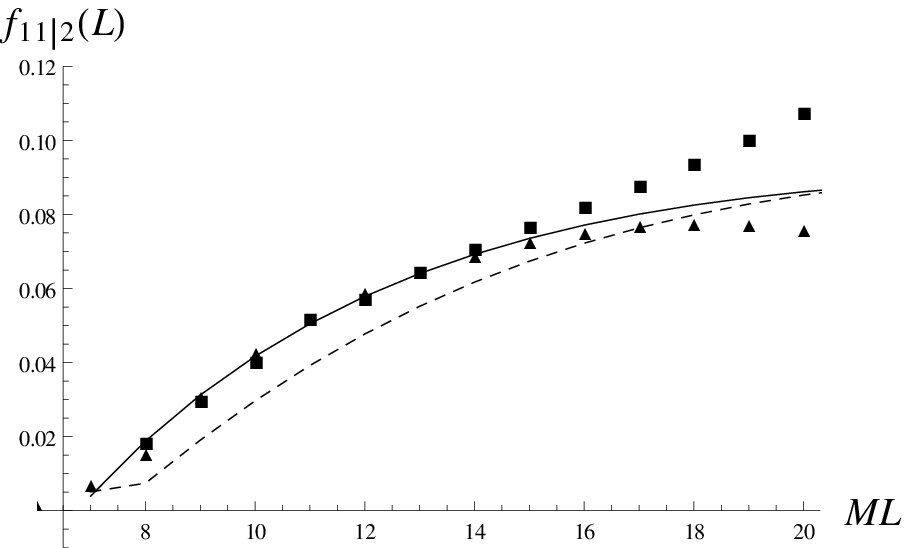} & \includegraphics[scale=0.85]{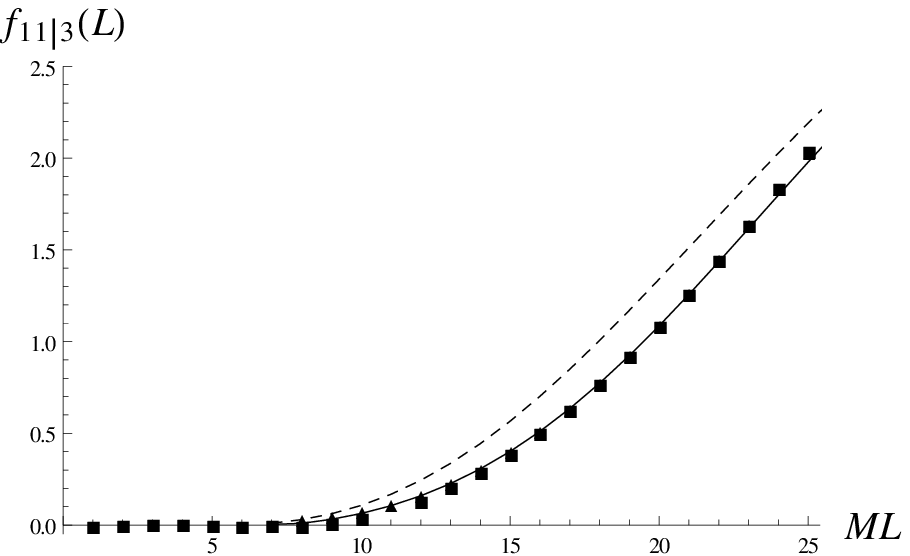}\tabularnewline
(a) $B_{2}$ & (b) $B_{3}$\tabularnewline
\end{tabular}
\par\end{centering}

\caption{\label{fig:twopartonepart}$\mu$-term corrections to the two-particle--one-particle
matrix element at $\xi=50/311$. The observable deviation for $L\gtrsim15$
in graph (a) is entirely due to numerical problems: the raw datum
determined from TCSA is the matrix element $\langle\{-\frac{1}{2},\frac{1}{2}\}|\mathcal{O}|\{0\}\rangle_{2,L}$,
which is so small in this range of the volume that it is of the same
magnitude as the truncation error itself. }
\end{figure}

Let us now study

\begin{equation}
f_{11|k}(L)=M^{-x}\sqrt{\rho_{11}(\tilde{\theta},-\tilde{\theta})}\sqrt{m_{k}L}\,\,_{11}\langle\{-\frac{1}{2},\frac{1}{2}\}|\mathcal{O}|\{0\}\rangle_{k,L}\label{eq:denscorr12}
\end{equation}
where $\tilde{\theta}$ solves the Bethe-Yang equation 
\[
m_{1}L\sinh\theta+\delta_{11}(2\theta)=\frac{1}{2}\quad,\quad\delta_{11}(\theta)=-i\log(-S_{11}(\theta))
\]
The expectation from the naive finite-volume formula (\ref{eq:genffrelation})
is that
\begin{equation}
f_{11|k}(L)=M^{-x}\left|F_{11k}^{\mathcal{O}}(i\pi-\tilde{\theta},i\pi+\tilde{\theta},0)\right|+O\left(\mathrm{e}^{-\mu L}\right)\label{eq:f2kas}
\end{equation}
where 
\[
F_{11k}^{\mathcal{O}}(\theta_{1},\theta_{2},\theta_{3})=\langle0|\mathcal{O}|B_{1}(\theta_{1})B_{1}(\theta_{2})B_{k}(\theta_{3})\rangle
\]
Once again, Fig. \ref{fig:twopartonepart} demonstrates that the exponential
corrections play an important role, and yield corrections of up to
20-30\%. This is very important because using the crossing property
(\ref{eq:crossing}) one can see that the matrix element measured
by the functions $f_{11|k}$ is 
\[
F_{11k}^{\mathcal{O}}(i\pi-\tilde{\theta},i\pi+\tilde{\theta},0)=\langle B_{1}(\tilde{\theta})B_{1}(-\tilde{\theta})|\mathcal{O}|B_{k}(\theta_{3})\rangle
\]
which for $k=3$ drives the decay $B_{3}\rightarrow B_{1}B_{1}$,
under a perturbation by the operator
\[
\frac{1}{2i}\left(\mathcal{O}-\mathcal{O}^{\dagger}\right)=\sin\frac{\beta}{2}\Phi
\]
Here we made use of the charge conjugation symmetry 
\[
\mathcal{C}:\qquad\Phi(x,t)\rightarrow-\Phi(x,t)
\]
of the model which ensures that 
\[
\langle B_{1}(\tilde{\theta})B_{1}(-\tilde{\theta})|\mathcal{O}^{\dagger}|B_{3}(\theta_{3})\rangle=-\langle B_{1}(\tilde{\theta})B_{1}(-\tilde{\theta})|\mathcal{O}|B_{3}(\theta_{3})\rangle
\]
because 
\[
\mathcal{C}\mathcal{O}\mathcal{C}^{-1}=\mathcal{O}^{\dagger}\mbox{ \,\,\ for\,\,\ }\mathcal{O}=\mbox{e}^{i\frac{\beta}{2}\Phi}
\]
and both $B_{1}$ and $B_{3}$ are odd under $\mathcal{C}$. According
to the data in Fig. \ref{fig:twopartonepart}, by omitting the $\mu$-term
contribution we would significantly underestimate the true infinite
volume matrix element (the dashed line) on the basis of the TCSA data.
Also note that although the decay of $\mu$-terms is exponential in
the asymptotic regime, in the volume range available in the numerical
method it could be so slow as to be practically unobservable as displayed
in Fig. \ref{fig:twopartonepart}. What is most surprising is the
marked contrast to the case of the energy level corrections in Fig.
\ref{fig:massmuterms}, which are driven by the same scales $\mu_{11}^{2}$
and $\mu_{12}^{3}$. 

It is shown in the next section that together with another source
of error (also related to the $\mu$-terms) this explains the dominant
source of systematic error encountered in the case of the $B_{3}$
resonance in the two-frequency sine-Gordon model \cite{Pozsgay:2006wb}.

\section{Decay widths of resonances in the two-frequency sine-Gordon model\label{sec:Decay-widths-of}}

Now we turn to an application of the results in the previous section.
Consider the two-frequency sine-Gordon model defined by the Hamiltonian

\begin{equation}
H_{DSG}=\int dx\,:\frac{1}{2}(\partial_{t}\Phi)^{2}+\frac{1}{2}(\partial_{x}\Phi)^{2}:-\int dx\left(\lambda:\cos\beta\Phi:+\zeta:\sin\frac{\beta}{2}\Phi:\right)\label{eq:dsg_ham}
\end{equation}
where normal-ordering is understood treating the field $\Phi$ as
a free massless boson, i.e. the theory is treated as a perturbed conformal
field theory similarly to (\ref{eq:pcft_action}). One can consider
this model to be a perturbation of sine-Gordon field theory by the
integrability-breaking coupling $\zeta$, and using the exact form
factors in the $\zeta=0$, a so-called form factor perturbation theory
(FFPT) can be developed \cite{Delfino:1996xp,Takacs:2009fu}. The
spectrum and phase structure of this model has been studied extensively
in the literature \cite{Delfino:1997ya,Nersesyan:2000,Bajnok:2000ar,Mussardo:2004rw,Takacs:2005fx,Takacs:2009fu}.
The integrability breaking interaction can be assigned the dimensionless
parameter 
\begin{equation}
t=\frac{\zeta}{M^{\frac{4+3\xi}{2+2\xi}}}\label{eq:DSG_dimless_coupling}
\end{equation}
This model can be considered as a toy model for hadrons: the $\zeta=0$
theory is in our setting analogous to QCD, with the breathers being
the mesons, which in this limit are all stable. Switching on $\zeta$
makes the mesons unstable, and the ones that are over the threshold
$2m_{1}$ are allowed to decay. Consider the decay of $B_{3}$ which
in the rest frame has the kinematics
\[
B_{3}(0)\rightarrow B_{1}(\theta_{c})+B_{1}(-\theta_{c})
\]
with the rapidities of the outgoing $B_{1}$s given by energy conservation.

\[
2\cosh(\theta_{c})=\frac{m_{3}}{m_{1}}=\frac{\sin\frac{3\pi\xi}{2}}{\sin\frac{\pi\xi}{2}}
\]
Using form factor perturbation theory, the decay width can be calculated
\cite{Pozsgay:2006wb}
\begin{equation}
\frac{\Gamma_{3\rightarrow11}}{M}=t^{2}\frac{s_{311}(\theta_{c})^{2}}{\left(\frac{m_{3}}{M}\right)^{2}\frac{m_{1}}{M}\sqrt{\left(\frac{m_{3}}{2m_{1}}\right)^{2}-1}}\label{eq:sg_decaywidth}
\end{equation}
where the (dimensionless) decay matrix element is given by
\[
s_{311}(\theta_{c})=\frac{f_{311}(\theta_{c})}{M^{\frac{\xi}{2+2\xi}}}\qquad\mbox{where}\qquad f_{311}(\theta_{c})=\left|F_{311}^{\Psi}\left(i\pi,\theta_{c},-\theta_{c}\right)\right|=\left|F_{311}^{\mathcal{O}}\left(i\pi,\theta_{c},-\theta_{c}\right)\right|
\]
where
\[
\Psi=\sin\frac{\beta}{2}\Phi(0)=\frac{1}{2i}(\mathcal{O}-\mathcal{O}^{\dagger})\qquad\mbox{with}\qquad\mathcal{O}=\mathrm{e}^{i\frac{\beta}{2}\Phi(0)}
\]
In the previous work \cite{Pozsgay:2006wb}, the prediction (\ref{eq:sg_decaywidth})
was compared to numerical TCSA data. We developed two methods, of
which we only consider the so-called ``improved mini-Hamiltonian''
(imH) method, because it is effectively a direct numerical determination
of the decay matrix element $s_{311}(\theta_{c})$ of the perturbing
Hamiltonian. 

The results are shown in Table \ref{cap:s311_measured} where we use
the following parameter for the sine-Gordon coupling:

\[
R=\frac{\sqrt{4\pi}}{\beta}\qquad\Rightarrow\qquad\xi=\frac{1}{2R^{2}-1}
\]
(the value $\xi=50/311$ used in section \ref{sec:Determining-matrix-elements}
corresponds to $R=1.9$).

\begin{table}
\begin{centering}
\begin{tabular}{|c|c|c|c|}
\hline 
$R$ & imH & FFPT & $ML_{0}$\tabularnewline
\hline 
\hline 
1.5 & $1.185\pm0.034$ & 1.1694 & 12.398\tabularnewline
\hline 
1.6 & $0.883\pm0.011$ & 0.9303 & 11.588\tabularnewline
\hline 
1.7 & $0.592\pm0.003$ & 0.6383 & 11.927\tabularnewline
\hline 
1.9 & $0.274\pm0.001$ & 0.2917 & 13.615\tabularnewline
\hline 
2.2 & $0.1046\pm0.0002$ & 0.0999 & 17.475\tabularnewline
\hline 
2.5 & $0.04767\pm2\cdot10^{-5}$ & 0.0392 & 22.309\tabularnewline
\hline 
2.6 & $0.03773\pm6\cdot10^{-5}$ & 0.0295 & 24.089\tabularnewline
\hline 
2.7 & $0.03022\pm5\cdot10^{-5}$ & 0.0224 & 25.946\tabularnewline
\hline 
\end{tabular}
\par\end{centering}

\caption{\label{cap:s311_measured} Measured values of $s_{311}$ using the
``improved mini-Hamiltonian'' method, compared to the theoretical
prediction labeled by FFPT. The last column contains the volume corresponding
to the level crossing at zero perturbing coupling. The uncertainties
shown are crude estimates of truncation errors from TCSA.}
\end{table}

Notice that the agreement is not particularly good. The errors are
estimates of the truncation errors coming from the energy cutoff introduced
in TCSA, which is analogous to the lattice spacing in QCD. For $R>1.7$
this clearly does not explain the deviation between the numerical
and theoretical results. It was already noted in the original paper
\cite{Pozsgay:2006wb} and it was (correctly) thought to originate
from exponential finite size corrections, however at that time the
tools to test this hypothesis had not yet been developed.

In subsection \ref{sub:Two-particle--one-particle-matri} we already
considered this particular matrix element in finite volume and noted
that the $\mu$-terms did explain the deviation between the naive
expectation (\ref{eq:genffrelation}) and the TCSA data. To investigate
whether this helps in the case of the resonance width determination,
we have to correct errors coming from two sources:
\begin{enumerate}
\item The position of the crossings between the $B_{3}$ and the $B_{1}B_{1}$
levels is not where one naively expects it to be. Neglecting $\mu$-terms,
this would be at the position where
\begin{equation}
E_{11}(L_{0})=m_{3}\label{eq:naivelceqn}
\end{equation}
with 
\begin{align*}
E_{11}(L) & =2m_{1}\cosh\tilde{\theta}
\end{align*}
where $\tilde{\theta}$ solves
\begin{equation}
m_{1}L\sinh\theta+\delta_{11}(\theta)=2\pi I\label{eq:b1b1by}
\end{equation}
with $I=1/2$ for the first two-particle level. \\
When incorporating the leading $\mu$-term correction to the $B_{3}$
level, the equation to solve is
\begin{equation}
E_{11}(L)=E_{3}(L)\label{eq:mulceqn}
\end{equation}
where
\[
E_{3}(L)=m_{1}(1+\cos2\tilde{u})
\]
with $\tilde{u}$ solving (\ref{eq:B3mutermby}) i.e.
\[
\mathrm{e}^{im_{1}L\sinh(\pm iu)}S_{11}(\pm iu)S_{11}(\pm2iu)=1
\]
The results are displayed in Table \ref{tab:Positions-of-line-crossings:}:
note that the $B_{3}$ $\mu$-term does explain the position of the
level crossing for $R\gtrsim1.9$ very well. The remaining discrepancy
results from further exponential corrections, both to the $B_{3}$
and the $B_{1}B_{1}$ level.\\
Note that as a result of the difference between (\ref{eq:naivelceqn})
and (\ref{eq:mulceqn}), the rapidity $\tilde{\theta}$ of the $B_{1}$
particles does not equal to $\theta_{c}$, so that means we are not
measuring the matrix element at the proper kinematical point. As a
result, the matrix element relevant to the comparison is 
\[
s_{311}(\theta_{1})
\]
 where $\theta_{1}$ is the solution of (\ref{eq:b1b1by}) at the
actual position of the line crossing.\\
\begin{table}
\begin{centering}
\begin{tabular}{|c|c|c|c|}
\hline 
$R$ & $ML_{0}$(TCSA) & $ML_{0}$(without $\mu$) & $ML_{0}$(with $\mu$)\tabularnewline
\hline 
\hline 
1.5 & 12.398 & 11.453 & 11.699\tabularnewline
\hline 
1.6 & 11.588 & 10.534 & 11.035\tabularnewline
\hline 
1.7 & 11.927 & 10.839 & 11.513\tabularnewline
\hline 
1.9 & 13.615 & 12.539 & 13.446\tabularnewline
\hline 
2.2 & 17.475 & 16.312 & 17.439\tabularnewline
\hline 
2.5 & 22.309 & 21.015 & 22.288\tabularnewline
\hline 
2.6 & 24.089 & 22.757 & 24.068\tabularnewline
\hline 
2.7 & 25.946 & 24.582 & 25.926\tabularnewline
\hline 
\end{tabular}
\par\end{centering}

\caption{\label{tab:Positions-of-line-crossings:} Positions of line-crossings:
comparing the measured values and the theoretical results obtained
without/with the $B_{3}$ $\mu$-corrections.}
\end{table}

\item The other effect comes from the $\mu$-term correction for the volume
dependence of the matrix element, which can be determined from the
third formula in (\ref{eq:b3ffmu}). Notice that this requires the
knowledge of the five-particle form factor function $F_{11111}$.
This can be used to correct the result obtained from the ``improved
mini-Hamiltonian'' method.
\end{enumerate}
The end result is summarized in Table \ref{cap:s311_corrected}. Notice
that the agreement between theory and numerics is improved by an order
of magnitude for $R\gtrsim1.9$, to a few percent deviation instead
of 10-50 \% (the almost exact equality at $R=1.9$ is simply a numerical
accident). 

\begin{table}
\begin{centering}
\begin{tabular}{|c|c|c|}
\hline 
$R$ & $\mu$-imH & $s_{311}(\theta_{1})$\tabularnewline
\hline 
\hline 
1.7 & $0.806(4)$ & 0.868\tabularnewline
\hline 
1.9 & $0.401$(1) & 0.401\tabularnewline
\hline 
2.2 & $0.138$ & 0.134\tabularnewline
\hline 
2.5 & $0.0528$ & 0.0509\tabularnewline
\hline 
2.6 & $0.0392$ & 0.0378\tabularnewline
\hline 
2.7 & $0.0294$ & 0.0285\tabularnewline
\hline 
\end{tabular}
\par\end{centering}

\caption{\label{cap:s311_corrected} Comparing the TCSA results for $s_{311}$
using the ``improved mini-Hamiltonian'' method with $\mu$-term
corrections, compared to the theoretical prediction taken at the modified
kinematical point $\theta_{1}$ instead of $\theta_{c}$.}
\end{table}

\section{Conclusions and outlook\label{sec:Conclusions-and-outlook}}

\subsection{Why are the $\mu$-terms so large?}

The first question to address is: how could the $\mu$-terms be so
large in the case of one-particle--one-particle and two-particle--one-particle
matrix elements? They decay exponentially by the volume, and therefore
are expected to decrease very quickly. As a rule of thumb volumes
larger than five times the correlation length of the theory are expected
to be large enough to safely neglect them when calculating to an accuracy
of a few percent. This is borne out by the finite volume mass corrections
examined in subsection \ref{sub:Finite-volume-mass}. It is important
to note that the $\mu$-term predictions for the mass corrections
read
\begin{align*}
E_{2}(L) & =2m_{1}\cos u_{2}\\
E_{3}(L) & =m_{1}(1+2\cos u_{3})
\end{align*}
with 
\begin{align*}
u_{2} & =\frac{\pi\xi}{2}+O(\mathrm{e}^{-\mu_{11}^{2}L})\\
u_{3} & =\pi\xi+O(\mathrm{e}^{-\mu_{12}^{3}L})
\end{align*}
Note that these enter as arguments of cosine functions, which are
very smooth and have derivatives of order $1$. As a result, their
contributions are relatively small for $\mu L>1$, with the asymptotic
forms 
\begin{align*}
E_{2}(L) & =m_{2}-\left(\gamma_{11}^{2}\right)^{2}\mu_{11}^{2}\mathrm{e}^{-\mu_{11}^{2}L}+\dots\\
E_{3}(L) & =m_{3}-2\left(\gamma_{12}^{3}\right)^{2}\mu_{12}^{3}\mathrm{e}^{-\mu_{12}^{3}L}+\dots
\end{align*}
What about matrix elements then? The great difference in the $\mu$-term
corrected finite-volume matrix elements (\ref{eq:vacb2ffmu},\ref{eq:b1b2ffmu}),
(\ref{eq:b3ffmu}) and (\ref{eq:b1b3ffimprmu}) is that the parameter
$u$ also appears in them as kinematical argument of form factors.
When the parameters $u$ are expanded around their asymptotic value,
the correction takes the form of the $\mu$-term exponential multiplied
by the derivative of the form factor%
\footnote{We have not written these expansions explicitly; the interested reader
can consult \cite{Pozsgay:2008bf} for examples.%
}. Form factors, however, can be rapidly varying functions of their
arguments. Indeed, they have singularities corresponding to bound
states and even in the absence of these they always have disconnected
contributions. These disconnected contributions can appear whenever
there are particle rapidities displaced by $i\pi$, which according
to the crossing relation (\ref{eq:crossing}) corresponds to a matrix
elements with one or more particles with identical quantum numbers
(including momenta) in the two states between which the matrix elements
of the local operator are taken.

The data are indeed consistent with this interpretation. As shown
on the example of vacuum--one-particle matrix elements, $\mu$-term
contributions are similar in magnitude to the mass corrections. There
are countless other data generated in previous studies (and also for
testing during this work) for matrix elements between vacuum and multi-particle
states, and the results are always similar: $\mu$-terms appear, but
their magnitude is consistent with what is expected from the energy
levels. They can only contribute much when their scale $\mu$ is very
small, i.e. for volumes $\mu L<1$ (which can still be a headache
in some cases, and this provided the motivation for the investigation
in \cite{Pozsgay:2008bf}).

However, for matrix elements with multi-particle states on both sides
one runs a good chance of running into problems with the rapid variation
of the form factor appearing in the $\mu$-term formulae. Sometimes
it is even possible to hit a disconnected piece head-on, as happened
for the stationary $B_{1}$-$B_{3}$ matrix element for which the
theoretical prediction (\ref{eq:b1b3ffimprmu}) did include an additional
term which introduced an additional factor of $L$ into the volume
dependence.

\subsection{Implications for the determination of matrix elements}

When matrix elements are to be determined numerically, whether in
lattice QCD or in the TCSA approach, it is necessary to model the
volume dependence of the matrix elements in order to extract the infinite
volume result one is interested in. The usual way to model it accounts
for the difference in the normalization between the infinite volume
and finite volume matrix elements, taken into account by the state
density factors in eqn. (\ref{eq:genffrelation}). It is important
to include the corrections by the scattering between the particles
in finite volume (taken into account by the Bethe-Yang equations (\ref{eq:breatherby})),
because these are of order of some inverse power of $L$. Indeed,
this is what Lellouch and Lüscher do in their work \cite{Lellouch:2000pv}.
Exponential finite size corrections, on the other hand, correspond
to radiative corrections specific to finite volume where the loop
winds around because of the periodic boundary conditions, and are
often neglected.

In some cases (notably those involving only stationary particles)
the properly normalized matrix elements are predicted to be constant
up to exponential finite size corrections. So the presence of such
corrections can be observed by the fact that the matrix element is
not a constant, as in subsections \ref{fig:vaconepart} and \ref{fig:onepartonepart}.
Therefore one could at least have the idea whether there are sizable
exponential corrections, and with a bit of luck one may even try to
fit it and extrapolate to infinite volume. However, in cases involving
moving particles as in subsection \ref{sub:Two-particle--one-particle-matri},
this does not work: there is no way of telling the qualitative difference
between the volume dependence of the dashed and continuous lines in
Fig. \ref{sub:Two-particle--one-particle-matri} without having a
theoretical understanding of the $\mu$-terms. 

If one tries to determine the infinite volume form factor without
accounting for the $\mu$-term, one can easily commit a (systematic)
error of the order of 20-50\% (for most of the parameter space in
the examples of two-particle--one-particle form factors considered
here), and sometimes even errors of the order of 100\% (for the $B_{1}-B_{2}$
case)!

Therefore the question is whether it is possible to have a theoretical
understanding of $\mu$-terms. In the case of integrable models the
job is made easier by the existence of exact solutions to the form
factor bootstrap, in terms of which one can express the $\mu$-term
contributions by analytically continuing the description (\ref{eq:genffrelationphasecorr}).
Note that for the case of $B_{1}B_{1}-B_{3}$ matrix elements, which
is of interest to resonances in the two-frequency sine-Gordon model,
this requires knowledge of a five-particle form factor, a quite nontrivial
object. Therefore this does not seem a viable option in the case of
theories like QCD, where the task is to determine weak interaction
matrix elements between hadron states.

We remark that $F$-term type exponential finite size corrections
are essentially perturbative when considered in terms of low energy
effective field theory; this is already clear from the derivation
of finite volume mass corrections by Lüscher \cite{Luscher:1985dn}.
In lattice QCD, they are handled by using chiral perturbation theory
to construct the loop integrals, which are then put into finite volume
using the finite volume perturbation theory formalism \cite{Gasser:1986vb,Gasser:1987ah,Gasser:1987zq,Sharpe:1992ft,Bernard:2001yj}.
The unknown matrix elements entering them are parametrized by some
low-energy constants, which are to be determined by fitting to volume
dependence numerically obtained on the lattice.

\subsection{Implications for resonances}

For resonances, there is not even the silver lining that existed for
matrix elements involving only stationary particles, where one could
at least observe if the volume dependence was not quite right. The
reason is that their properties can be measured by observing level
splittings, which occur at particular values of the volume. Even in
the two-dimensional setting, finding two or more level crossings (more
properly these are level avoidances) of the same type (meaning the
same particle content for the levels) at substantially different values
of the volume, and with the numerics accurate enough, is a tough problem
\cite{Pozsgay:2006wb}. However, to see whether there are substantial
exponential finite size corrections to the decay rates one needs to
do exactly that: the most straightforward way out is to find level
crossings at several different values of the volume, and perform an
infinite volume extrapolation, or at least get an estimate for the
systematic error caused by the exponential corrections.

\subsection{Closing remarks}

Without having a handle on the size of the exponential finite size
corrections one cannot tell whether the observed discrepancy e.g.
between a lattice result and an experimentally measured decay width
is a result of systematic error or a disagreement between theoretical
model and physical reality. Therefore it is important to have a good
theoretical description, or at least an estimate of these corrections.
One special feature of $\mu$-terms is that in contrast to all other
exponential corrections which are bounded by the value of the mass
gap of the theory, $\mu$-terms can decay substantially slower with
the volume in the case of weakly bound states. Moreover, even when
they are expected to decay fast on the basis of the magnitude of their
characteristic scale $\mu$, their contributions to matrix elements
can still be enhanced. Measuring their contribution to energy levels
is not enough to estimate the errors in matrix element and resonance
width calculations.

It is important to stress that the ingredients essential to the observations
made in this papers are not restricted to the two-dimensional setting.
The description of particles as composites in subsection \ref{sub:mu-terms-the-leading-exp-correction}
is just the well-known quantum mechanical treatment of bound states
as analytic continuation of scattering states to complex values of
the momenta. In fact, the expression of $\mu$-term corrections to
finite volume masses are not that much different in higher space-time
dimensions \cite{Luscher:1985dn}, the only difference is the appearance
of some (negative) powers of $L$ (and numerical factors) in front
of the exponential correction term. Concerning form factors, the matrix
elements in quantum field theories of four (or indeed any) space-time
dimensions share the essential feature with their two-dimensional
counterparts that they have singularities corresponding to disconnected
pieces and bound state poles%
\footnote{The only special feature in two-dimensions is that so-called anomalous
threshold singularities which are branch cuts in four dimensions,
correspond to poles in two-dimensional theories \cite{Coleman:1978kk}. %
}. Therefore they too can be expected to show rapid variation in kinematical
domains close to these singularities. In fact, the difficult aspect
of modeling the $\mu$-terms in higher dimensions is to have the necessary
information about many-particle form factors, which is so readily
provided by the bootstrap in integrable models.

\subsubsection*{Acknowledgments}

GT is grateful to B. Pozsgay and S. Katz for discussions. This work
was partially supported by the Hungarian OTKA grants K75172 and K81461.

\appendix
\makeatletter 
\renewcommand{\theequation}{\hbox{\normalsize\Alph{section}.\arabic{equation}}} \@addtoreset{equation}{section} \renewcommand{\thefigure}{\hbox{\normalsize\Alph{section}.\arabic{figure}}} \@addtoreset{figure}{section} \renewcommand{\thetable}{\hbox{\normalsize\Alph{section}.\arabic{table}}} \@addtoreset{table}{section} \makeatother

\section{\label{sec:Sine-Gordon-breather-form-factors}Sine-Gordon breather
form factors}

\subsection{Spectrum and $S$ matrix\label{sub:Spectrum-and-}}

The fundamental excitations of the sine-Gordon model (\ref{eq:sg_action})
are a doublet of soliton/antisoliton of mass $M$. Their exact $S$
matrix can be written as \cite{zam-zam} 
\begin{equation}
\mathcal{S}_{i_{1}i_{2}}^{j_{1}j_{2}}(\theta,\xi)=S_{i_{1}i_{2}}^{j_{1}j_{2}}(\theta,\xi)S_{0}(\theta,\xi)\label{eq:sg_smatrix}
\end{equation}
where
\begin{eqnarray*}
 &  & S_{++}^{++}(\theta,\xi)=S_{--}^{--}(\theta,\xi)=1\\
 &  & S_{+-}^{+-}(\theta,\xi)=S_{-+}^{-+}(\theta,\xi)=S_{T}(\theta,\xi)\\
 &  & S_{+-}^{-+}(\theta,\xi)=S_{-+}^{+-}(\theta,\xi)=S_{R}(\theta,\xi)
\end{eqnarray*}
and
\begin{eqnarray*}
 &  & S_{T}(\theta,\xi)=\frac{\sinh\left(\frac{\theta}{\xi}\right)}{\sinh\left(\frac{i\pi-\theta}{\xi}\right)}\qquad,\qquad S_{R}(\theta,\xi)=\frac{i\sin\left(\frac{\pi}{\xi}\right)}{\sinh\left(\frac{i\pi-\theta}{\xi}\right)}\\
 &  & S_{0}(\theta,\xi)=-\exp\left\{ -i\int_{0}^{\infty}\frac{dt}{t}\frac{\sinh\frac{\pi(1-\xi)t}{2}}{\sinh\frac{\pi\xi t}{2}\cosh\frac{\pi2}{2}}\sin\theta t\right\} \\
 &  & =-\left(\prod_{k=1}^{n}\frac{ik\pi\xi-\theta}{ik\pi\xi+\theta}\right)\exp\Bigg\{-i\int_{0}^{\infty}\frac{dt}{t}\sin\theta t\\
 &  & \quad\times\frac{\left[2\sinh\frac{\pi(1-\xi)t}{2}\mathrm{e}^{-n\pi\xi t}+\left(\mathrm{e}^{-n\pi\xi t}-1\right)\left(\mathrm{e}^{\pi(1-\xi)t/2}+\mathrm{e}^{-\pi(1+\xi)t/2}\right)\right]}{2\sinh\frac{\pi\xi t}{2}\cosh\frac{\pi2}{2}}\Bigg\}
\end{eqnarray*}
(the latter representation is valid for any value of $n\in\mathbb{N}$
and makes the integral representation converge faster and further
away from the real $\theta$ axis). 

Besides the solitons, the spectrum of theory contains also breathers
$B_{r}$, with masses
\begin{equation}
m_{r}=2M\sin\frac{r\pi\xi}{2}\label{eq:breather_mass}
\end{equation}
The breather-soliton and breather-breather $S$-matrices are also
explicitly known \cite{zam-zam}.

\subsection{Breather form factors}

We only consider exponentials of the bosonic field $\Phi$. To obtain
matrix elements containing the first breather, one can analytically
continue the elementary form factors of sinh-Gordon theory obtained
in \cite{Koubek:1993ke} to imaginary values of the couplings. The
result is
\begin{eqnarray}
F_{\underbrace{{\scriptstyle 11\dots1}}_{n}}^{a}(\theta_{1},\dots,\theta_{n}) & = & \left\langle 0\left|\mathrm{e}^{ia\beta\Phi(0)}\right|B_{1}(\theta_{1})\dots B_{1}(\theta_{n})\right\rangle \nonumber \\
 & = & \mathcal{G}_{a}(\beta)\,[a]_{\xi}\,(i\bar{\lambda}(\xi))^{n}\,\prod_{i<j}\frac{f_{\xi}(\theta_{j}-\theta_{i})}{\mathrm{e}^{\theta_{i}}+\mathrm{e}^{\theta_{j}}}\, Q_{a}^{(n)}\left(\mathrm{e}^{\theta_{1}},\dots,\mathrm{e}^{\theta_{n}}\right)\label{eq:b1ffs}
\end{eqnarray}
where 
\begin{eqnarray*}
Q_{a}^{(n)}(x_{1},\dots,x_{n}) & = & \det{[a+i-j]_{\xi}\,\sigma_{2i-j}^{(n)}(x_{1},\dots,x_{n})}_{i,j=1,\dots,n-1}\mbox{ if }n\geq2\\
Q_{a}^{(1)} & = & 1\quad,\qquad[a]_{\xi}=\frac{\sin\pi\xi a}{\sin\pi\xi}\\
\bar{\lambda}(\xi) & = & 2\cos\frac{\pi\xi}{2}\sqrt{2\sin\frac{\pi\xi}{2}}\exp\left(-\int_{0}^{\pi\xi}\frac{dt}{2\pi}\frac{t}{\sin t}\right)
\end{eqnarray*}
and 
\begin{eqnarray}
f_{\xi}(\theta) & = & v(i\pi+\theta,-1)v(i\pi+\theta,-\xi)v(i\pi+\theta,1+\xi)\nonumber \\
 &  & v(-i\pi-\theta,-1)v(-i\pi-\theta,-\xi)v(-i\pi-\theta,1+\xi)\nonumber \\
v(\theta,\zeta) & = & \prod_{k=1}^{N}\left(\frac{\theta+i\pi(2k+\zeta)}{\theta+i\pi(2k-\text{\ensuremath{\zeta}})}\right)^{k}\exp\Bigg\{\int_{0}^{\infty}\frac{dt}{t}\Big(-\frac{\zeta}{4\sinh\frac{t}{2}}-\frac{i\text{\ensuremath{\zeta}}\theta}{2\pi\cosh\frac{t}{2}}\nonumber \\
 &  & +\left(N+1-N\mbox{e}^{-2t}\right)\mbox{e}^{-2Nt+\frac{it\theta}{\pi}}\frac{\sinh\zeta t}{2\sinh^{2}t}\Big)\Bigg\}\label{eq:brminff}
\end{eqnarray}
gives the minimal $B_{1}B_{1}$ form factor%
\footnote{The formula for the function $v$ is in fact independent of $N$;
choosing $N$ large extends the width of the strip where the integral
converges and also speeds up convergence.%
}, while $\sigma_{k}^{(n)}$ denotes the elementary symmetric polynomial
of $n$ variables and order $k$ defined by
\[
\prod_{i=1}^{n}(x+x_{i})=\sum_{k=0}^{n}x^{n-k}\sigma_{k}^{(n)}(x_{1},\dots,x_{n})
\]
Furthermore
\begin{eqnarray}
\mathcal{G}_{a}(\beta)=\langle e^{ia\beta\Phi}\rangle & = & \left[\frac{M\sqrt{\pi}\Gamma\left(\frac{4\pi}{8\pi-\beta^{2}}\right)}{2\Gamma\left(\frac{\beta^{2}/2}{8\pi-\beta^{2}}\right)}\right]^{\frac{a^{2}\beta^{2}}{4\pi}}\exp\Bigg\{\int_{0}^{\infty}\frac{dt}{t}\Bigg[-\frac{a^{2}\beta^{2}}{4\pi}e^{-2t}\nonumber \\
 &  & +\frac{\sinh^{2}\left(\frac{a}{4\pi}t\right)}{2\sinh\left(\frac{\beta^{2}}{8\pi}t\right)\cosh\left(\left(1-\frac{\beta^{2}}{8\pi}\right)t\right)\sinh t}\Bigg]\Bigg\}\label{eq:exactvev}
\end{eqnarray}
is the exact vacuum expectation value of the exponential field \cite{Lukyanov:1996jj},
with $M$ denoting the soliton mass related to the coupling $\lambda$
in (\ref{eq:pcft_action}) via \cite{Zamolodchikov:1995xk} 
\begin{equation}
\lambda=\frac{2\Gamma(\Delta)}{\pi\Gamma(1-\Delta)}\left(\frac{\sqrt{\pi}\Gamma\left(\frac{1}{2-2\Delta}\right)M}{2\Gamma\left(\frac{\Delta}{2-2\Delta}\right)}\right)^{2-2\Delta}\qquad,\qquad\Delta=\frac{\beta^{2}}{8\pi}\label{eq:mass_scale}
\end{equation}
Formula (\ref{eq:b1ffs}) also coincides with the result given in
\cite{Lukyanov:1997bp}. Form factors containing higher breathers
can be obtained using that $B_{n}$ is a bound state of $B_{1}$ and
$B_{n-1}$; therefore sequentially fusing $n$ adjacent first breathers
gives $B_{n}$. Following the lines of reasoning of Appendix A of
the paper \cite{Pozsgay:2006wb} one obtains 
\begin{eqnarray}
 &  & F_{k_{1}\dots k_{r}nl_{1}\dots l_{s}}^{a}(\theta_{1},\dots,\theta_{r},\theta,\theta_{1}',\dots,\theta_{s}')=\label{eq:higherbff}\\
 &  & \left\langle 0\left|\mathrm{e}^{ia\beta\Phi(0)}\right|B_{k_{1}}(\theta_{1})\dots B_{k_{r}}(\theta_{r})B_{n}(\theta)B_{l_{1}}(\theta_{1}')\dots B_{l_{s}}(\theta_{s}')\right\rangle =\gamma_{11}^{2}\gamma_{12}^{3}\dots\gamma_{1n-1}^{n}\nonumber \\
 &  & \times F_{k_{1}\dots k_{r}\underbrace{{\scriptstyle 11\dots1}}_{n}l_{1}\dots l_{s}}^{a}\left(\theta_{1},\dots,\theta_{r},\theta+\frac{1-n}{2}i\pi\xi,\theta+\frac{3-n}{2}i\pi\xi,\dots,\theta+\frac{n-1}{2}i\pi\xi,\theta_{1}',\dots,\theta_{s}'\right)\nonumber 
\end{eqnarray}
where
\begin{equation}
\gamma_{1k}^{k+1}=\sqrt{\frac{2\tan\frac{k\pi\xi}{2}\tan\frac{(k+1)\pi\xi}{2}}{\tan\frac{\pi\xi}{2}}}\label{eq:gammacoeffs}
\end{equation}
is the $B_{1}B_{k}\rightarrow B_{k+1}$ coupling, defined as the residue
of the appropriate scattering amplitude:
\begin{eqnarray}
i\left(\gamma_{1k}^{k+1}\right)^{2} & = & \mathop{\mbox{Res}}_{\theta=\frac{i\pi(k+1)\xi}{2}}S_{1k}(\theta)\nonumber \\
S_{1k}(\theta) & = & \frac{\sinh\theta+i\sin\frac{\pi(k+1)\xi}{2}}{\sinh\theta-i\sin\frac{\pi(k+1)\xi}{2}}\frac{\sinh\theta+i\sin\frac{\pi(k-1)\xi}{2}}{\sinh\theta-i\sin\frac{\pi(k-1)\xi}{2}}\label{eq:b1bksmat}
\end{eqnarray}

\providecommand{\href}[2]{#2}\begingroup\raggedright\endgroup


\begin{thebibliography}{10}

\bibitem{Maiani:1990ca}
L.~Maiani and M.~Testa, ``{Final state interactions from Euclidean correlation
  functions},''
\href{http://dx.doi.org/10.1016/0370-2693(90)90695-3}{{\em Phys. Lett.}
  {\bfseries B245} (1990) 585--590}.

\bibitem{Lellouch:2000pv}
L.~Lellouch and M.~Luscher, ``{Weak transition matrix elements from
  finite-volume correlation functions},''
  \href{http://dx.doi.org/10.1007/s002200100410}{{\em Commun. Math. Phys.}
  {\bfseries 219} (2001) 31--44},
\href{http://arxiv.org/abs/hep-lat/0003023}{{\ttfamily arXiv:hep-lat/0003023
  [hep-lat]}}.

\bibitem{Lin:2001ek}
C.~Lin, G.~Martinelli, C.~T. Sachrajda, and M.~Testa, ``{$K\rightarrow \pi \pi$
  decays in a finite volume},''
  \href{http://dx.doi.org/10.1016/S0550-3213(01)00495-3}{{\em Nucl. Phys.}
  {\bfseries B619} (2001) 467--498},
  \href{http://arxiv.org/abs/hep-lat/0104006}{{\ttfamily arXiv:hep-lat/0104006
  [hep-lat]}}.

\bibitem{Luscher:1991cf}
M.~Luscher, ``{Signatures of unstable particles in finite volume},''
  \href{http://dx.doi.org/10.1016/0550-3213(91)90584-K}{{\em Nucl. Phys.}
  {\bfseries B364} (1991) 237--254}.

\bibitem{zam-zam}
A.~B. Zamolodchikov and A.~B. Zamolodchikov, ``{Factorized S-matrices in two
  dimensions as the exact solutions of certain relativistic quantum field
  models},''
\href{http://dx.doi.org/10.1016/0003-4916(79)90391-9}{{\em Annals Phys.}
  {\bfseries 120} (1979) 253--291}.

\bibitem{Mussardo:1992uc}
G.~Mussardo, ``{Off critical statistical models: Factorized scattering theories
  and bootstrap program},''
\href{http://dx.doi.org/10.1016/0370-1573(92)90047-4}{{\em Phys. Rept.}
  {\bfseries 218} (1992) 215--379}.

\bibitem{Karowski:1978vz}
M.~Karowski and P.~Weisz, ``{Exact Form-Factors in (1+1)-Dimensional Field
  Theoretic Models with Soliton Behavior},''
\href{http://dx.doi.org/10.1016/0550-3213(78)90362-0}{{\em Nucl. Phys.}
  {\bfseries B139} (1978) 455}.

\bibitem{Kirillov:1987jp}
A.~N. Kirillov and F.~A. Smirnov, ``{A representation of the current algebra
  connected with the SU(2) invariant Thirring model},''
\href{http://dx.doi.org/10.1016/0370-2693(87)90908-7}{{\em Phys. Lett.}
  {\bfseries B198} (1987) 506--510}.

\bibitem{Smirnov:1992vz}
F.~A. Smirnov, ``{Form-factors in completely integrable models of quantum field
  theory},''
{\em Adv. Ser. Math. Phys.} {\bfseries 14} (1992) 1--208.

\bibitem{Yurov:1989yu}
V.~P. Yurov and A.~B. Zamolodchikov, ``{Truncated conformal space approach to
  scaling Lee-Yang model},''
\href{http://dx.doi.org/10.1142/S0217751X9000218X}{{\em Int. J. Mod. Phys.}
  {\bfseries A5} (1990) 3221--3246}.

\bibitem{Pozsgay:2006wb}
B.~Pozsgay and G.~Takacs, ``{Characterization of resonances using finite size
  effects},'' \href{http://dx.doi.org/10.1016/j.nuclphysb.2006.05.007}{{\em
  Nucl. Phys.} {\bfseries B748} (2006) 485--523},
\href{http://arxiv.org/abs/hep-th/0604022}{{\ttfamily arXiv:hep-th/0604022}}.

\bibitem{Pozsgay:2007kn}
B.~Pozsgay and G.~Takacs, ``{Form factors in finite volume I: form factor
  bootstrap and truncated conformal space},''
  \href{http://dx.doi.org/10.1016/j.nuclphysb.2007.06.027}{{\em Nucl. Phys.}
  {\bfseries B788} (2008) 167--208},
\href{http://arxiv.org/abs/0706.1445}{{\ttfamily arXiv:0706.1445}}.

\bibitem{Pozsgay:2007gx}
B.~Pozsgay and G.~Takacs, ``{Form factors in finite volume II:disconnected
  terms and finite temperature correlators},''
  \href{http://dx.doi.org/10.1016/j.nuclphysb.2007.07.008}{{\em Nucl. Phys.}
  {\bfseries B788} (2008) 209--251},
\href{http://arxiv.org/abs/0706.3605}{{\ttfamily arXiv:0706.3605}}.

\bibitem{Pozsgay:2008bf}
B.~Pozsgay, ``{Luscher's mu-term and finite volume bootstrap principle for
  scattering states and form factors},''
  \href{http://dx.doi.org/10.1016/j.nuclphysb.2008.04.021}{{\em Nucl. Phys.}
  {\bfseries B802} (2008) 435--457},
\href{http://arxiv.org/abs/0803.4445}{{\ttfamily arXiv:0803.4445 [hep-th]}}.

\bibitem{Bajnok:2000wm}
Z.~Bajnok, L.~Palla, G.~Takacs, and F.~Wagner, ``{The k-folded sine-Gordon
  model in finite volume},''
  \href{http://dx.doi.org/10.1016/S0550-3213(00)00441-7}{{\em Nucl. Phys.}
  {\bfseries B587} (2000) 585--618},
\href{http://arxiv.org/abs/hep-th/0004181}{{\ttfamily arXiv:hep-th/0004181}}.

\bibitem{Feher:2011aa}
G.~Feher and G.~Takacs, ``{Sine-Gordon form factors in finite volume},''
  \href{http://dx.doi.org/10.1016/j.nuclphysb.2011.06.020}{{\em Nucl. Phys.}
  {\bfseries B852} (2011) 441--467},
  \href{http://arxiv.org/abs/1106.1901}{{\ttfamily arXiv:1106.1901 [hep-th]}}.

\bibitem{Luscher:1986pf}
M.~Luscher, ``{Volume Dependence of the Energy Spectrum in Massive Quantum
  Field Theories. 2. Scattering States},''
  \href{http://dx.doi.org/10.1007/BF01211097}{{\em Commun. Math. Phys.}
  {\bfseries 105} (1986) 153--188}.

\bibitem{Luscher:1990ux}
M.~Luscher, ``{Two particle states on a torus and their relation to the
  scattering matrix},''
  \href{http://dx.doi.org/10.1016/0550-3213(91)90366-6}{{\em Nucl. Phys.}
  {\bfseries B354} (1991) 531--578}.

\bibitem{Luscher:1985dn}
M.~Luscher, ``{Volume Dependence of the Energy Spectrum in Massive Quantum
  Field Theories. 1. Stable Particle States},''
\href{http://dx.doi.org/10.1007/BF01211589}{{\em Commun. Math. Phys.}
  {\bfseries 104} (1986) 177}.

\bibitem{Klassen:1990ub}
T.~R. Klassen and E.~Melzer, ``{On the relation between scattering amplitudes
  and finite size mass corrections in QFT},''
  \href{http://dx.doi.org/10.1016/0550-3213(91)90566-G}{{\em Nucl. Phys.}
  {\bfseries B362} (1991) 329--388}.

\bibitem{Feverati:1998va}
G.~Feverati, F.~Ravanini, and G.~Takacs, ``{Truncated conformal space at c = 1,
  nonlinear integral equation and quantization rules for multi-soliton
  states},'' \href{http://dx.doi.org/10.1016/S0370-2693(98)00543-7}{{\em Phys.
  Lett.} {\bfseries B430} (1998) 264--273},
\href{http://arxiv.org/abs/hep-th/9803104}{{\ttfamily arXiv:hep-th/9803104}}.

\bibitem{Delfino:1996xp}
G.~Delfino, G.~Mussardo, and P.~Simonetti, ``{Non-integrable Quantum Field
  Theories as Perturbations of Certain Integrable Models},''
  \href{http://dx.doi.org/10.1016/0550-3213(96)00265-9}{{\em Nucl. Phys.}
  {\bfseries B473} (1996) 469--508},
\href{http://arxiv.org/abs/hep-th/9603011}{{\ttfamily arXiv:hep-th/9603011}}.

\bibitem{Takacs:2009fu}
G.~Takacs, ``{Form factor perturbation theory from finite volume},''
  \href{http://dx.doi.org/10.1016/j.nuclphysb.2009.10.001}{{\em Nucl. Phys.}
  {\bfseries B825} (2010) 466--481},
\href{http://arxiv.org/abs/0907.2109}{{\ttfamily arXiv:0907.2109}}.

\bibitem{Delfino:1997ya}
G.~Delfino and G.~Mussardo, ``{Nonintegrable aspects of the multifrequency
  Sine-Gordon model},''
  \href{http://dx.doi.org/10.1016/S0550-3213(98)00063-7}{{\em Nucl. Phys.}
  {\bfseries B516} (1998) 675--703},
  \href{http://arxiv.org/abs/hep-th/9709028}{{\ttfamily arXiv:hep-th/9709028
  [hep-th]}}.

\bibitem{Nersesyan:2000}
M.~Fabrizio, A.~O. Gogolin, and A.~A. Nersesyan, ``{Critical properties of the
  double-frequency sine-Gordon model with applications},''
  \href{http://dx.doi.org/10.1016/S0550-3213(00)00247-9}{{\em Nucl. Phys.}
  {\bfseries B580} (2000) 647--687},
  \href{http://arxiv.org/abs/cond-mat/0001227}{{\ttfamily
  arXiv:cond-mat/0001227 [cond-mat]}}.

\bibitem{Bajnok:2000ar}
Z.~Bajnok, L.~Palla, G.~Takacs, and F.~Wagner, ``{Nonperturbative study of the
  two frequency sine-Gordon model},''
  \href{http://dx.doi.org/10.1016/S0550-3213(01)00067-0}{{\em Nucl. Phys.}
  {\bfseries B601} (2001) 503--538},
\href{http://arxiv.org/abs/hep-th/0008066}{{\ttfamily arXiv:hep-th/0008066}}.

\bibitem{Mussardo:2004rw}
G.~Mussardo, V.~Riva, and G.~Sotkov, ``{Semiclassical particle spectrum of
  double Sine-Gordon model},''
  \href{http://dx.doi.org/10.1016/j.nuclphysb.2004.04.003}{{\em Nucl. Phys.}
  {\bfseries B687} (2004) 189--219},
  \href{http://arxiv.org/abs/hep-th/0402179}{{\ttfamily arXiv:hep-th/0402179
  [hep-th]}}.

\bibitem{Takacs:2005fx}
G.~Takacs and F.~Wagner, ``{Double sine-Gordon model revisited},''
  \href{http://dx.doi.org/10.1016/j.nuclphysb.2006.02.004}{{\em Nucl. Phys.}
  {\bfseries B741} (2006) 353--367},
\href{http://arxiv.org/abs/hep-th/0512265}{{\ttfamily arXiv:hep-th/0512265}}.

\bibitem{Gasser:1986vb}
J.~Gasser and H.~Leutwyler, ``{Light Quarks at Low Temperatures},''
  \href{http://dx.doi.org/10.1016/0370-2693(87)90492-8}{{\em Phys. Lett.}
  {\bfseries B184} (1987) 83}.

\bibitem{Gasser:1987ah}
J.~Gasser and H.~Leutwyler, ``{Thermodynamics of Chiral Symmetry},''
  \href{http://dx.doi.org/10.1016/0370-2693(87)91652-2}{{\em Phys. Lett.}
  {\bfseries B188} (1987) 477}.

\bibitem{Gasser:1987zq}
J.~Gasser and H.~Leutwyler, ``{Spontaneously Broken Symmetries: Effective
  Lagrangians at Finite Volume},''
  \href{http://dx.doi.org/10.1016/0550-3213(88)90107-1}{{\em Nucl. Phys.}
  {\bfseries B307} (1988) 763}.

\bibitem{Sharpe:1992ft}
S.~R. Sharpe, ``{Quenched chiral logarithms},''
  \href{http://dx.doi.org/10.1103/PhysRevD.46.3146}{{\em Phys. Rev.} {\bfseries
  D46} (1992) 3146--3168},
  \href{http://arxiv.org/abs/hep-lat/9205020}{{\ttfamily arXiv:hep-lat/9205020
  [hep-lat]}}.

\bibitem{Bernard:2001yj}
{\bfseries MILC} Collaboration, C.~Bernard, ``{Chiral logs in the presence of
  staggered flavor symmetry breaking},''
  \href{http://dx.doi.org/10.1103/PhysRevD.65.054031}{{\em Phys. Rev.}
  {\bfseries D65} (2002) 054031},
  \href{http://arxiv.org/abs/hep-lat/0111051}{{\ttfamily arXiv:hep-lat/0111051
  [hep-lat]}}.

\bibitem{Coleman:1978kk}
S.~R. Coleman and H.~Thun, ``{On the prosaic origin of the double poles in the
  sine-Gordon S matrix},'' \href{http://dx.doi.org/10.1007/BF01609466}{{\em
  Commun. Math. Phys.} {\bfseries 61} (1978) 31}.

\bibitem{Koubek:1993ke}
A.~Koubek and G.~Mussardo, ``{On the operator content of the sinh-Gordon
  model},'' \href{http://dx.doi.org/10.1016/0370-2693(93)90554-U}{{\em Phys.
  Lett.} {\bfseries B311} (1993) 193--201},
\href{http://arxiv.org/abs/hep-th/9306044}{{\ttfamily arXiv:hep-th/9306044}}.

\bibitem{Lukyanov:1996jj}
S.~L. Lukyanov and A.~B. Zamolodchikov, ``{Exact expectation values of local
  fields in quantum sine-Gordon model},''
  \href{http://dx.doi.org/10.1016/S0550-3213(97)00123-5}{{\em Nucl. Phys.}
  {\bfseries B493} (1997) 571--587},
\href{http://arxiv.org/abs/hep-th/9611238}{{\ttfamily arXiv:hep-th/9611238}}.

\bibitem{Zamolodchikov:1995xk}
A.~B. Zamolodchikov, ``{Mass scale in the sine-Gordon model and its
  reductions},''
\href{http://dx.doi.org/10.1142/S0217751X9500053X}{{\em Int. J. Mod. Phys.}
  {\bfseries A10} (1995) 1125--1150}.

\bibitem{Lukyanov:1997bp}
S.~L. Lukyanov, ``{Form factors of exponential fields in the sine-Gordon
  model},'' \href{http://dx.doi.org/10.1142/S0217732397002673}{{\em Mod. Phys.
  Lett.} {\bfseries A12} (1997) 2543--2550},
\href{http://arxiv.org/abs/hep-th/9703190}{{\ttfamily arXiv:hep-th/9703190}}.

\end{thebibliography}
\end{document}